\documentclass[a4paper,fleqn,usenatbib]{mnras}

\usepackage{newtxtext,newtxmath}
\usepackage[T1]{fontenc}
\usepackage{ae,aecompl}

\usepackage{graphicx}	
\usepackage{amsmath}	
\usepackage{amssymb}	
\usepackage{cleveref}

\usepackage{subfig}

\title{ALMA observations of lensed Herschel sources : Testing the dark-matter halo paradigm}

\author[A. Amvrosiadis et al.]{
A. Amvrosiadis$^{1}$\thanks{E-mail: AmvrosiadisA@cardiff.ac.uk},
S. A. Eales$^{1}$,
M. Negrello$^{1}$,
L. Marchetti$^{2, 3}$,
M. W. L. Smith$^{1}$,   \newauthor
N. Bourne$^{4}$,
D. L. Clements$^{5}$,
G. De Zotti$^{6}$,
L. Dunne$^{1}$,
S. Dye$^{7}$,
C. Furlanetto$^{1}$,  \newauthor
R. J. Ivison$^{4, 8}$
S. Maddox$^{1}$,
E. Valiante$^{1}$,
M. Baes$^{9}$,
A. J. Baker$^{10}$,
A. Cooray$^{11}$,  \newauthor
S. M. Crawford$^{12}$,
D. Frayer$^{13}$,
A. Harris$^{14}$,
M. J. Micha{\l}owski$^{4, 15}$,
H. Nayyeri$^{11}$, \newauthor
S. Oliver$^{16}$,
D. A. Riechers$^{17}$,
S. Serjeant$^{3}$,
M. Vaccari$^{18, 19}$
\\
\\
$^{1}$School of Physics and Astronomy, Cardiff University, The Parade, Cardiff CF24 3AA, UK \\
$^{2}$Department of Physics and Astronomy, University of the Western Cape, Private Bag X17, 7535, Bellville, Cape Town, South Africa \\
$^{3}$School of Physical Sciences, The Open University, Walton Hall, Milton Keynes MK7 6AA, UK \\
$^{4}$Institute for Astronomy, University of Edinburgh, Royal Observatory, Edinburgh EH9 3HJ, UK \\
$^{5}$Astrophysics Group, Imperial College, Blackett Laboratory, Prince Consort Road, London SW7 2AZ, UK \\
$^{6}$INAF-Osservatorio Astronomico di Padova, Vicolo dell Osservatorio 5, I-35122 Padova, Italy \\
$^{7}$School of Physics and Astronomy, University of Nottingham, University Park, Nottingham NG7 2RD, UK \\
$^{8}$Department of Physical Science, The Open University, Milton Keynes MK7 6AA, UK \\
$^{9}$Sterrenkundig Observatorium, Universiteit Gent, Krijgslaan 281 S9, B-9000 Gent, Belgium \\
$^{10}$Department of Physics and Astronomy, Rutgers, the State University of New Jersey, 136 Frelinghuysen Road, Piscataway, NJ 08854-8019, USA \\
$^{11}$Department of Physics and Astronomy, University of California, Irvine, CA 92697, USA \\
$^{12}$South African Astronomical Observatory, PO Box 9, Observatory, 7935 Cape Town, South Africa \\
$^{13}$National Radio Astronomy Observatory, Green Bank, WV 24944, USA \\
$^{14}$Department of Astronomy University of Maryland College Park, MD 20742, USA \\
$^{15}$Astronomical Observatory Institute, Faculty of Physics, Adam Mickiewicz University, ul.~S{\l}oneczna 36, 60-286 Pozna{\'n}, Poland \\
$^{16}$Department of Physics and Astronomy, University of Sussex, Brighton BN1 9QH, UK \\
$^{17}$Department of Astronomy, Cornell University, Ithaca, NY 14853, USA \\
$^{18}$Department of Physics and Astronomy, University of the Western Cape, Robert Sobukwe Road, 7535 Bellville, Cape Town, South Africa \\
$^{19}$INAF Istituto di Radioastronomia, via Gobetti 101, I-40129 Bologna, Italy \\
}

\pubyear{2017}

\begin{document}
\label{firstpage}
\pagerange{\pageref{firstpage}--\pageref{lastpage}}
\maketitle

\begin{abstract}

With the advent of wide-area submillimeter surveys, a large number of high-redshift gravitationally lensed dusty star-forming galaxies (DSFGs) has been revealed. Due to the simplicity of the selection criteria for candidate lensed sources in such surveys, identified as those with $S_{500\mu m} > 100$ mJy, uncertainties associated with the modelling of the selection function are expunged. The combination of these attributes makes submillimeter surveys ideal for the study of strong lens statistics. We carried out a pilot study of the lensing statistics of submillimetre-selected sources by making observations with the Atacama Large Millimetre Array (ALMA) of a sample of strongly-lensed sources selected from surveys carried out with the Herschel Space Observatory. We attempted to reproduce the distribution of image separations for the lensed sources using a halo mass function taken from a numerical simulation which contains both dark matter and baryons. We used three different density distributions, one based on analytical fits to the halos formed in the EAGLE simulation and two density distributions (Singular Isothermal Sphere (SIS) and SISSA) that have been used before in lensing studies. We found that we could reproduce the observed distribution with all three density distributions, as long as we imposed an upper mass transition of $\sim$$10^{13} M_{\odot}$ for the SIS and SISSA models, above which we assumed that the density distribution could be represented by an NFW profile. We show that we would need a sample of $\sim$500 lensed sources to distinguish between the density distributions, which is practical given the predicted number of lensed sources in the Herschel surveys.

\end{abstract}

\begin{keywords}
galaxies : high-redshift - gravitational lensing: strong - submillimetre : galaxies
\end{keywords}

\section{Introduction}  \label{sec:section_1}

Photons traveling from a distant background source and
through the vicinity of massive objects, such as galaxies or
groups/clusters of galaxies, get deflected by the presence of their gravitational
field.
If the background source and the foreground object are well-aligned with the observer, we have the creation of multiple images. This effect is called strong gravitational lensing \citep{SchneiderEhlersFalco1992}.

For a sample of these lensed sources, the statistics of angular separations
mainly depend on four factors: (a) the luminosity function of the source population \citep{More2012};
(b) the number-density of dark-matter halos
as a function of halo mass and redshift \citep{Eales2015}; (c) the mass density distributions
within the halos \citep{TakahashiChiba2001, KochanekWhite2001, Oguri2002b, Oguri2006b}; (d) the cosmological model \citep{LiOstriker2002, LiOstriker2003, Chae2003, Oguri2008, Oguri2012}.
In principle, therefore, the
statistics of image separations for a suitable sample of lensed
sources is a powerful way of examining the mass density distribution of the total matter in the halo and halo mass functions predicted by simulations.

The two alternative methods for producing samples of strong lenses
for statistical purposes are to start from either
a population of objects that potentially act as
lenses or from a population of potentially lensed sources.
Follow-up observations are necessary in both cases to confirm
the strong lensing nature.
Examples of the first method are
the Sloan Lens ACS (SLACS) Survey
\citep{Bolton2006}
and the BOSS Emission-Line Lens Survey (BELLS) \citep{Brownstein2012}, in
both of which
the potential lensed systems were found by looking for galaxies
with a spectrum which show two redshifts - with confirmation of
the lensing provided by imaging with the Hubble Space Telescope.
For our purpose of investigating the properties of halos, the disadvantage of this approach is that it is prone to selection effects.

Examples of the second method were the
Cosmic Lens All-Sky Survey (CLASS) \citep{Myers2003,Browne2003}
and the Sloan Digital Sky Surveys Quasar Lens Search (SQLS) \citep{Oguri2006a}.
CLASS was the
the largest survey of strongly lensed quasars conducted
at radio wavelengths.
Starting from a well-defined statistical sample
of
$\sim 9000$ flat-spectrum radio sources, the CLASS team used high-resolution
radio observations to produce a statistically well-defined sample
of 13 lensed sources \citep{Browne2003}.
The SQLS selected potential lens candidates from the Sloan Digital Sky Survey
\citep{Oguri2006a}, producing a final
catalogue
\citep{Inada2012} of 26 lensed quasars
from an initial catalogue of
$\sim50000$ quasars. It is worth pointing out that both optical and radio surveys require huge parent samples in order to identify a few strong lenses.

With the advent of wide-area extragalactic surveys undertaken
with {\it Herschel Space Observatory} \citep{Pilbratt2010} at
submillimeter wavelengths on the other hand, a new method for discovering high-redshift
gravitationally lensed dusty star-forming
galaxies has been made possible with an almost 100 per cent efficiency. The number counts of
un-lensed submillimeter galaxies (SMGs) are very steep at
bright flux densities \citep{Blain1996, Negrello2007}. Therefore, the brightest sources
after removing nearby galaxies and radio-loud AGN can be selected as candidate lensed sources,
since there are very few high-redshift galaxies that are intrinsicly so
bright (effectively exploiting an extreme case of the magnification bias). Negrello et al.
(2010) demonstrated  this method for the first time,
using the initial results from
the {\it Herschel} Astrophysical Terahertz Large Area Survey (H-ATLAS; Eales et al. 2012) .
They showed that
out of ten extragalactic sources with flux $S > 100$ mJy at 500 $\mu$m,
five were
strongly lensed high-redshift galaxies, with the remainder being
easily identified as local $(z<0.1)$ spiral galaxies and in one case a
previously known radio-bright AGN. Exploiting the whole $\sim 600 \,$ deg$^2$ area
covered by H-ATLAS, Negrello et al. (2017) have identified a
sample of 80 candidate strongly lensed SMGs using the same
selection criteria. Follow-up observations with submillimetre interferometers
or with the {\it Hubble Space Telescope} and {\it W. M. Keck Observatory}
have confirmed so far that 20 of
these extragalactic sources show a strong lensing morphology.
Samples of lensed sources have now been produced using the
same method from other
{\it Herschel} surveys.
A sample of 13 candidate strongly
lensed galaxies was produced by Wardlow et
al. (2013) from $95 \,$ deg$^2$ of
the {\it Herschel} Multi-tiered Extragalactic Survey (HerMES; Oliver et al. 2012) , 11 of which have been confirmed by follow-up
observations to be strongly lensed. More recently,
Nayeri et al. (2016) produced a list of 77 candidate
gravitationally lensed galaxies from the HerMES Large Mode
Survey (HeLMS; Oliver et al. 2012) and the
{\it Herschel} Stripe 82 Survey (HerS; Viero et al. 2014) , which in total cover an area of $372 \,$ deg$^2$.

This uniform
and simple selection technique, which identifies potential candidates based on the
emission from the source rather than the lens and so
falls in the second category of methods.
One of the main advantages of this technique is that sub-millimeter emission
from the
lens is usually negligible compared with the
emission from the source.
Therefore, the modelling of the lensed source emission in high
resolution submillimeter/millimeter imaging data does not
suffer from uncertainties caused by the lens subtraction \citep{Dye2014, Negrello2014}.

Bussmann et al. (2013; B13)
presented
Sub-Millimeter Array (SMA) 880 $\mu$m observations of a sample of 30 candidates strong gravitational lenses identified from the
two widest {\it Herschel} extragalactic surveys,
H-ATLAS \& HerMES.
In our previous paper (Eales 2015) we investigated whether the standard dark-matter halo
paradigm could explain the distribution of Einstein radii measured
from the SMA observations.
We tried three halo mass functions, all estimated from numerical simulations that only
included dark matter, and two different methods for calculating the
lensing magnification produced by each dark-matter halo. In all cases we found that the
model predicted a larger number of sources with large Einstein radii than we
observed.
In this paper, we have extended and improved our previous study in
several ways. First, the SMA results we used in our previous paper
had limited angular resolution and sensitivity, and we were concerned that
we might have missed arcs of large angular size
 with low surface brightness, causing
us to underestimate the number of sources with large image separations.
For this reason, we started a project to map the lensed Herschel sources
with ALMA, and in this paper we present the first results from this
ALMA project. We compare the distributions of image separations measured from
the ALMA images with the predictions of our models.
Our second improvement is to use a halo mass function and density
distributions from the halos derived
from a numerical simulation that includes baryons as well as dark
matter.

The layout
of this paper is as follows.
In Section 2, we present the first results from
our ALMA project.
In Section 3
we describe the halo models and lay down the theoretical background
for computing the lensing properties of the halos.
Section 4 describes the comparison between the observed and predicted Einstein
radii. We discuss our results in Section 5.
Throughout this paper, we assume
a flat $\Lambda$CDM model with the best-fit parameters derived
from the results from the {\it Planck Observatory} \citep{Planck2014}, which are $\Omega_m = 0.307$ and $h=0.693$.

\section{The Pilot Sample and the ALMA Observations}  \label{sec:section_2}

ALMA has much better angular resolution and surface-brightness sensitivity
than the SMA, making it a much better instrument for mapping a strongly-lensed
submillimetre source.
In our previous SMA study of the lensing statistics
of strongly-lensed Herschel sources (B13), the limited angular resolution
of the SMA
meant that it was often not clear whether the structure seen on the maps
was
actually due to lensing. There is also the
possibility that
large arcs were missed by their falling below
the surface-brightness limit of the SMA.
Since the new ALMA observations would be so much
better than the SMA observations, we defined
a new sample of sources for our ALMA programme.

Negrello et al. (2010) showed that it is possible to select a sample of lensed
sources from a Herschel survey with close to 100\% efficiency.
Of the Herschel sources with 500-$\mu$m flux densities > 100 mJy, roughly
one half are strongly lensed and half are a mixture of nearby galaxies
and radio-loud AGN. Negrello et al. (2010) showed that it is actually
very easy removing these contaminants, since nearby galaxies are
easy to identify by inspecting optical surveys, such as the Sloan Digital
Sky Survey, and radio-loud AGN are easy to spot because they are found
in radio surveys. After rejecting contaminants in this way, Negrello
et al. (2010) achieved a 100\% success rate for their initial sample.

A number of Herschel teams have used this method to select samples
of sources that are probably lensed and then used molecular-line spectroscopy to
measure redshifts for the sources.
A slight variant on the basic method followed by most of
these teams is to use the ratios of fluxes in the Herschel bands
to select sources that are likely to have redshifts in the wavelength
range covered by their spectrometer
(e.g. Harris et al. 2012; Lupu et al. 2012), which will create a
slight bias towards certain redshift ranges.

An accurate lensing model for a source requires it
to have an accurate
redshift. Therefore, as the initial sample for our ALMA programme,
we selected
42 sources from
the H-ATLAS and HELMS surveys
with the highest 500-$\mu$m flux densities
and
with spectroscopic redshifts $>1$.
We checked that none of our candidates is a radio-loud AGN.
In almost all cases, the 500-$\mu$m flux densities of the
sources are >100 mJy, the flux limit used by Negrello et al.
(2010). The lower redshift limit, of course,
removes any nearby galaxies, and so
we expect virtually all of the sources
to be strongly lensed. For the reasons described above,
the requirement that the sources have spectroscopic redshifts
has probably introduced a slight bias towards certain redshift ranges,
but the conditional
probability statistics we use in this paper (see Section~\ref{sec:section_4}) ensure that
our results will not be affected by this bias.
Of the 42 sources, only
16 were finally observed by ALMA before the end of Cycle 2, but this should not introduce any bias because we did not rank the sources in priority. Table~\ref{tab:ALMA_DATA} lists the sample of 16 sources.

\begin{figure*}
\includegraphics[width=0.85\textwidth, height=0.95\textheight]{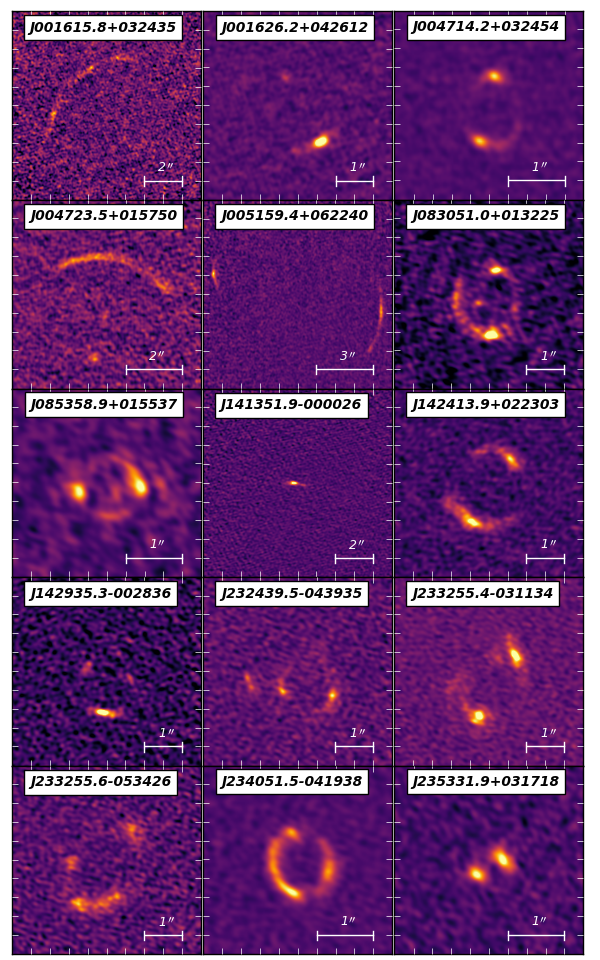}
\caption{The 873-$\mu$m continuum emission images of the 15 sources we observed with ALMA. 
The source HATLAS J083344.9+000109, which was part of the observing run, has been neglected because it doesn't reveal any lensing features. The flux axes are not shown on the same scale for all the lens systems, as the large arcs would appear very faint. North is up and East is left}
\label{fig:ALMA_IMAGES}
\end{figure*}

We observed each source for approximately 2-4 minutes
with ALMA at 873 $\mu$m with a maximum baseline of 1km,
which gives an angular resolution of 0.12 arcsec. The final image products were produced by the standard ALMA pipeline. The lensed sources are shown in Figure~\ref{fig:ALMA_IMAGES}, all exept one. The source HATLAS J083344.9+000109 is barely detected in the ALMA image and is the faintest 500-$\mu$m source in the sample. There are no obvious signs of lensing features, either on the ALMA image or on the optical image from the Sloan Digital Sky Survey. This source is coincident with a QSO. In addition, the source HATLAS J141351.9-000026 does not seem to have any lensing structure. However, as seen from Figure 3 in Negrello et al. 2017 the ALMA image captures a small part of large faint arc.

\begin{table*}
	\caption{The ALMA sample}
	\centering{
	\begin{tabular}{llccccc}
		\hline
		\hline
		IAU Name & Other  & 500-$\mu$m flux & $z_l$ &  $z_s$ &  $\theta_E \, ['']$ & Ref. \\
                         & name   & density [mJy] & & & & \\
		\hline
HeLMS J001615.8+032435 & HeLMS13 & 149$\pm$7  & 0.663 & 2.765 & 5.22 $\pm$ 0.05 & N16 \\
HeLMS J001626.2+042612 & HeLMS22 & 127$\pm$7  & 0.2154 & 2.509 & 0.98 $\pm$ 0.05 & M17, N16 \\
HeLMS J004714.2+032453 & HeLMS8 & 168$\pm$8  & 0.478 & 1.195 &  0.58 $\pm$ 0.05 & N16 \\
HeLMS J004723.5+015750 & HeLMS9 & 164$\pm$8  & 0.3650 &  1.441 &  2.66 $\pm$ 0.05 & M17, N16 \\
HeLMS J005159.4+062240 & HeLMS18 & 135$\pm$7 &  - & 2.392 &  6.54 $\pm$ 0.05 & N16\\
H-ATLAS J083051.0+013225 & G09v1.97& 269$\pm$9 & 0.626  &  3.634 & 0.85 $\pm$ 0.04 & B13, MN17 \\
H-ATLAS J083344.9+000109 & - & 96$\pm$9 & - & 2.530 & - & M17 \\
H-ATLAS J085358.9+015537 & G09v1.40 & 228$\pm$9 & -  &  2.089	 & 0.55$\pm$0.04 & B13, MN16, M17 \\
H-ATLAS J141351.9-000026 & G15v2.235 & 176$\pm$9 & 0.547 & 2.478 & -   & B13, H12, MN17 \\
H-ATLAS J142413.9+022303 & G15v2.779 & 193$\pm$9 & 0.595 & 4.243 &  1.02 $\pm$ 0.04  & C11, B13, MN17  \\
H-ATLAS J142935.3-002836  & G15v2.19 & 200$\pm$8 & 0.218 &  1.027 &  0.71 $\pm$ 0.04    & C14, M14, MN17 \\
HeLMS J232439.5-043935 & HeLMS7 & 172$\pm$9 & - & 2.473 & 0.65 $\pm$ 0.05 & N16 \\
HeLMS J233255.4-031134 & HeLMS2 & 263$\pm$8  & 0.426 &  2.689 & 0.93 $\pm$ 0.05  & N16 \\
HeLMS J233255.6-053426 & HeLMS15 & 147$\pm$9  & 0.976 &  2.402  & 0.98 $\pm$ 0.05 & N16\\
HeLMS J234051.5-041938 & HeLMS5 & 205$\pm$8  &  - & 3.503 &  0.54 $\pm$ 0.05 & N16\\
HeLMS J235331.9+031718 & HeLMS40 & 111$\pm$7 & 0.821 &  - & 0.26 $\pm$ 0.05  &N16\\
		\hline
		\hline
	\end{tabular}
	\\
	}
	\begin{flushleft}
	\textbf{Notes:} Column $\theta_E$ corresponds to the Einstein radius, which is half the image separation. The references, from which the lens and source redshift were obtained, are as follows:
C11 =  \citep{Cox2011}; B13 = \citep{Bussmann2013}; C14 = \citep{Calanog2014};
N16 = \citep{Nayyeri2016}; MN17 = \citep{Negrello2017}; M17 = Marchetti et al. in prep.
	\end{flushleft}
	\label{tab:ALMA_DATA}
\end{table*}

For the remaining sources in the sample, there is clear evidence of strong lensing features in the ALMA images. Modelling of the submillimetre emission, by constructing detailed lensing models, will be presented in two upcoming papers (Dye et al. in preparation; Negrello et al. in preparation). The Einstein radii were measured directly from the images and subsequently compared with the respective values that arise from preliminary lensing models of these system, whereupon an agreement was confirmed. In cases where only an arc is visible (e.g HeLMS J001615.8+032435) a rough estimate of the Einstein radius was performed by fitting a circle to the peaks of the emission $( > 4\sigma)$. A uniform weighting was applied to these pixels, alleviating any dependence on their fluxes and taking into account only on their positions.

For three sources (H-ATLAS J083051.0+013225, H-ATLAS J085358.9+015537, H-ATLAS J142413.9+022303) there are also measurements of the Einstein radius from SMA observations (see Table~\ref{tab:SMA_DATA}). For these sources, the pairs of measurements, with the SMA measurement first are: $0.39 \pm 0.02$ and $0.85 \pm 0.04$ arcsec; $0.55 \pm 0.04$ and $0.55 \pm 0.04$ arcsec; $0.57 \pm 0.01$ and $1.02 \pm 0.04$ arcsec.
This disagreement in the inferred values of the Einstein radii can be attributed to the complex structure of the submillimeter emission which can not be fully resolved with the SMA observations, as well as the complexity of the foreground mass distribution \citep{Bussmann2013}.

\begin{table*}
	\caption{The SMA sample}
	\centering{
	\begin{tabular}{llrrcc}
		\hline
		\hline
		IAU Name & Name & $z_l$ &  $z_s$ &  $\theta_E \, ['']$ & Ref. \\
		\hline
H-ATLAS J083051.0+013225 & G09v1.97  	 & 0.6260   &  3.6340 &  0.39 $\pm$ 0.02   & B13  \\
H-ATLAS J085358.9+015537 & G09v1.40  	 & -  &  2.0894& 0.55 $\pm$ 0.04 & B13, M17 \\
H-ATLAS J090302.9-014127  & SDP17      	 & 0.9435 & 2.3049  &   0.33 $\pm$ 0.02   & N10, B13 \\
H-ATLAS J090311.6+003906 & SDP81      	 & 0.2999 &  3.0420 	& 1.52 $\pm$ 0.03   & N10, B13 \\
H-ATLAS J090740.0-004200  & SDP9      	 & 0.6129 & 1.5770 	 &   0.59 $\pm$ 0.04   & N10, B13 \\
H-ATLAS J091043.1-000321  & SDP11      	 & 0.7930 &  1.7860	 &  0.95 $\pm$ 0.02   & N10, B13 \\
H-ATLAS J091305.0-005343  & SDP130     	& 0.2201 &  2.6260 	 &  0.43 $\pm$ 0.07   & N10, B13 \\
H-ATLAS J114637.9-001132   & G12v2.30  	& 1.2247  & 3.2590  &   0.65 $\pm$ 0.02   & B13, O13 \\
H-ATLAS J125135.4+261457  & NCv1.268 	& - & 3.6750   & 1.02 $\pm$ 0.03   &B13\\
H-ATLAS J125632.7+233625  & NCv1.143 	& 0.2551 &  3.5650 &    0.68 $\pm$ 0.01   &B13\\
H-ATLAS J132427.0+284449  & NBv1.43   	& 0.9970  & 1.6760  & -   & G05, G13\\
H-ATLAS J132630.1+334410 & NAv1.195  	& 0.7856 &  2.9510 & 1.80 $\pm$ 0.02   &B13\\
H-ATLAS J133649.9+291801 & NAv1.144 	& - &  2.2024 & 0.40 $\pm$ 0.03   & B13, O13 \\
H-ATLAS J133542.9+300401 & - 	 & 0.980 &  2.6850 & - &S14, R17 \\
H-ATLAS J133846.5+255054 & -  & 0.420 	&  2.4900 & - & N17 \\
H-ATLAS J134429.4+303036 & NAv1.56 & 0.6721 &  2.3010 &  0.92 $\pm$ 0.02   &H12, B13 \\
H-ATLAS J142413.9+022303 & G15v2.779 	& 0.5950 &  4.243 &  0.57 $\pm$ 0.01   & B13  \\
HERMES J021830.5-053124 & HXMM02  	& 1.350 &  3.3950 &  0.44 $\pm$ 0.02 & B13, W13    \\
HERMES J105712.2+565457 & HLock03 	& - &  2.7710 &  - & W13 \\
HERMES J105750.9+573026 & HLock01	& 0.600 & 2.9560 &  3.86 $\pm$ 0.01 & B13, W13  \\
HERMES J110016.3+571736 & HLock12 & 0.630 & 1.6510 &  1.14 $\pm$ 0.04 & C14 \\
HERMES   J142825.5+345547 & HBootes02 	& 0.414 & 2.8040 &  0.77 $\pm$ 0.03    &B13, W13  \\
		\hline
		\hline
	\end{tabular}
	\\
	}
	\begin{flushleft}
	\textbf{Note:} Column $\theta_E$ corresponds to the Einstein radius, which is half the image separation. The references, from which the lens and source redshift were obtained as well as the estimates for the Einstein radii,
are as follows: G05 = \citep{GladdersYee2005};
N10 = \citep{Negrello2010}; H12 = \citep{Harris2012}; B13 = \citep{Bussmann2013}; G13 = \citep{George2013};
O13 = \citep{Omont2013};  W13 = \citep{Wardlow2013}; C14 = \citep{Calanog2014}; D14 = \citep{Dye2014};
M14 = \citep{Messias2014}; S14 = \citep{Stanford2014}; N16 = \citep{Nayyeri2016};
M17 = Marchetti et al. in prep.; R17 = Riechers et al. in prep.
	\end{flushleft}
	\label{tab:SMA_DATA}
\end{table*}


\section{Methodology}  \label{sec:section_3}

In this section we describe the methodology for predicting the distribution of image separations. In section~\ref{sec:subsection_3.1} we discuss the different density profiles that were considered in this work. In section~\ref{sec:subsection_3.2} we present the halo mass function model. In section~\ref{sec:subsection_3.3} we describe the standard approach for computing lensing properties assuming spherical symmetry and finally in section~\ref{sec:subsection_3.4} we lay down the formalism for computing strong lensing statistics.

\subsection{The Halo Density Profiles}  \label{sec:subsection_3.1}

In the dark matter halo paradigm, galaxies are forming in an evolving population of dark matter haloes. High-resolution pure dark matter N-body simulations have been used extensively to study this dark component of the universe. These studies suggest that the spatial mass density distribution of dark matter, inside the halos identified in simulations, is well fitted by a single profile across a wide range of halo masses, the NFW profile \citep{Navarro1996, Navarro1997}. The NFW density profile is given by
\begin{equation}
\rho(r) = \frac{\rho_s}{(r/r_s)(1 + r/r_s)^2}
\end{equation}
where $r_s=r_{\textit{vir}}/c$ is the scale radius with $c$ being the concentration parameter which is approximated by the formula
\begin{equation}
\label{eq:concentration_Prada2011}
c (M,z)= 5 \left( \frac{M_{\textit{h}}}{10^{13} M_{\odot}} \right)^{-0.074} \left( \frac{1 + z}{1.7} \right)^{-1}
\end{equation}
and is derived from numerical simulations of Prada et al. 2011.

However, the objects that we observe in the real universe are comprised of both dark and baryonic matter. The difficulty is in producing density profiles for halos that also include baryons, because the physics of how baryons accrete into the centre of the halo and the astrophysical processes that take place in these central regions are complex and poorly understood. Two different analytic approaches are considered in this study, in an attempt to describe the total mass density distribution in early-type galaxies.

The simplest approach, which is frequently used in the literature is the Singular Isothermal Sphere (SIS) model. The SIS density profile is given by
\begin{equation}
\rho(r) = \frac{\sigma_v^2}{2\pi G r^2} \, ,
\end{equation}
where $G$ is the gravitational constant and $\sigma_v$ is the velocity dispersion of the halo. The later can be determined from the circular velocity of the halo, $V^2 = G M_{\textit{h}}/r_{\textit{vir}}$,

following the commonly used assumption that $\sigma_v \approx V/\sqrt{2}$.There are strong observational evidence that this power law model provides a good description of the total mass distribution in field early-type galaxies. Joint gravitational lensing and stellar-dynamical analysis of a sample of strong lenses from the SLACS survey, does indeed confirm that the average logarithmic slope for the total mass density is $\langle \gamma \rangle \simeq 2.0$ with some intrinsic scatter \citep{Koopmans2006, Koopmans2009}. Similar analysis was performed for the first five strong gravitational lens systems discovered in H-ATLAS \citep{Dye2014}, where the results found were in agreement with previous studies.

Recently, Lapi et al. (2012) adopted a rather theoretical approach by considering the contribution from baryons and dark matter, separately. They used an NFW profile to represent the mass density distribution for the dark matter component and a Sersic profile for the stellar component. The three-dimensional functional form of the Sersic profile \citep{PrugnielSimien1997} is given by,
\begin{equation}
\rho(r) = \frac{M_{\star}}{4 \pi R_e^3}\frac{b_n^{2n}}{n\Gamma(2n)} \left( \frac{r}{R_e} \right)^{-\alpha_n} exp\left[ -b_n \left( \frac{r}{R_e} \right)^{1/n}\right] \, ,
\end{equation}
where $n$ is the Sersic index, $R_e$ is the effective radius, $b_n = 2n - 1/3 + 0.009876/n$, $a_n = 1 - 1.188/2n + 0.22/4n^2$ and $M_{\star}$ is the stellar mass. The stellar mass can be determined by assuming a fixed ratio between the halo and stellar mass $M_h / M_{\star}$.

Lapi et al. (2012) showed that for galaxy-scale lenses this model, hereafter referred to as the SISSA model, yields very similar results to the SIS model under the assumption of reasonable parameters. However, this model has two additional free parameters that are affected by a large scatter. The first parameter is the ratio of halo to stellar mass, which for early-type galaxies is expected to lie in the range of $10-70$. The second parameter is the concentration parameter, $c$, which is expected to have a $20\%$ scatter.
In our analysis we will omit the scatter in the c-M relation and adopt a constant ratio of halo to stellar mass of 30. However, we show in Appendix~\ref{sec:Appendix_A} how these parameters can affect our results.

An additional parameter that is introduced in the above mentioned models is the virialization redshift $z_{l,v}$. This parameter is used to determine the virial radius of the halo $r_{\textit{vir}}$
\begin{equation}
r_{\textit{vir}} = \left( \frac{3 M_{\textit{h}}}{4 \pi \Delta_c \rho_{\textit{crit}}} \right)^{1/3} \, ,
\end{equation}
where $\rho_{\textit{crit}}(z) = \rho_{\textit{crit},0}E^2(z)$ is the critical density of the universe at redshift z, with $\rho_{\textit{crit},0}$ being it's value at redshift zero and $E(z)$ is the scaled Hubble parameter,
\begin{equation}
E^2(z) = \frac{H^2(z)}{H_0^2} =  \Omega_{m,0}(1+z)^3 + \Omega_{\Lambda, 0}(1+z)^{3(1+w)} \, .
\end{equation}
Assuming a flat cosmology ($\Omega_m + \Omega_{\Lambda} = 1$) we can use an approximate expression for $\Delta_c$, which was derived from a fit to simulations of \cite{BryanNorman1998},
\begin{equation}
\Delta_c = 18\pi^2 + 82x  - 39x^2 \, ,
\end{equation}
where $x = \Omega_m(z) - 1$ and the redshift evolution of the cosmological parameter of matter is
\begin{equation}
\Omega_m(z) = \frac{\rho_m}{\rho_{\textit{crit}}} =  \Omega_{m,0}(1+z)^3/E^2(z) \, .
\end{equation}

Lapi et al. (2012) suggested that the frequently made approximation, that the observed redshift of a galaxy is equal to the virialization redshift $z_l \approx z_{l,v}$, leads to an overestimation of the halo size. Alternatively they propose a virialization redshift in the range $z_{l,v} \sim 1.5-3.5$, which is much more in line with the ages of the stellar populations found in early-type galaxies.

Besides the analytic models presented above, we also now have results from cosmological hydrodynamic simulations which provide the means to examine how baryonic effects modify the structure of dark matter halos in a more rigorous way.  In recent studies, Schaller et al. 2015a,b investigated the internal structure of halos produced in the EAGLE simulations, which include both baryons and dark matter \citep{Schaye2015}. Some of the baryonic effects that are included in these simulation runs are feedback processed from massive stars and active galactic nuclei (AGN), radiative cooling, and contraction of the dark matter in the central halo regions due to the presence of baryons. The authors demonstrated that the following formula,
\begin{equation} \label{eq:EAGLE}
\frac{\rho(r)}{\rho_{\textit{crit}}} = \frac{\delta_s}{\left( r/r_s \right) \left( 1 + r/r_s \right)^2} + \frac{\delta_i}{\left( r/r_i \right) \left( 1 + \left(r/r_i \right)^2\right)} \, ,
\end{equation}
provides a good fit to the data. From the above functional form we clearly see that the first term is the NFW profile which provides a fairly good description of the outer part of the halo. The second term is an NFW-like profile with a steeper slope to account for the concentration of baryons in the central region of the halo. The parameters of this model as a function of mass, namely $\delta_s$, $r_s$, $\delta_i$ and $r_i$, are determined by fitting 3rd-order polynomials to the values found in Table 2 of Schaller et al. 2015a. The halo mass range probed in this study ranges from $M_h = 10^{10} - 10^{14} \, M_{\odot}$.

\subsection{Halo Mass Function}  \label{sec:subsection_3.2}

The halo mass function describes the comoving number density of dark matter halos as a function of redshift and per comoving mass interval. In our earlier paper \citep{Eales2015}, we used analytic functions, obtained by fitting to the results of numerical simulations of the evolution of dark matter, of Sheth \& Tormen (1999; ST99) and Tinker et al. (2008; T08). We found very little difference between the results predicted from the two halo mass functions. Both these analytic functions were based on numerical simulations containing only dark matter. In this paper, we use the analytic function for the halo mass function that was derived by Bocquet et al. (2016) by fitting to the results of a numerical simulation that contains both baryons and dark matter, using the same formalism as T08. The comoving number density of haloes of mass M is given by
\begin{equation}
	\frac{dn}{dM} = f(\sigma) \frac{\bar{\rho}_m}{M} \frac{dln\sigma^{-1}}{dM}
\end{equation}
where $\bar{\rho}_m$ is the mean number density at the current epoch and $\sigma$ is the square root of the variance of the mass-density field
\begin{equation}
	\sigma^2 = \left\langle \, \left( \frac{\delta M}{M} \right)^2 \right\rangle = \frac{1}{2 \pi^2} \int P^{\textit{lin}}(k,z) \hat{W}^2(kR)k^2dk,
\end{equation}
which is smoothed on a scale of comoving radius $R=( 3 M / 4 \pi \bar{\rho}_{m,0})^{1/3}$, using the Fourier transform of the real-space top-hat filter,
\begin{equation}
	\hat{W}(kR) = 3 \frac{sin(kR) - (kR) cos(kR)}{(kR)^3} \, .
\end{equation}
The function $f(\sigma)$ is parametrized as
\begin{equation}
	f(\sigma) = A \left[ \left( \frac{\sigma}{b} \right)^{-\alpha} + 1 \right] e^{-c / \sigma^2}
\end{equation}
where the parameters $A$, $\alpha$, $b$ and $c$ are all expressed as functions of redshift $A(z) = A_0(1+z)^{A_z}$, $\alpha(z) = \alpha_0(1+z)^{\alpha_z}$, $b(z) = b_0(1+z)^{b_z}$ and $c(z) = c_0(1+z)^{c_z}$. The best fit values of these parameters are obtained from Table 2 of Bocquet et al. (2016) for the Hydro simulation.

\subsection{Lensing Properties}  \label{sec:subsection_3.3}

In our analysis we consider the typical lensing configuration which is comprised of a point-like source located at redshift $z_s$, an object acting as a lens located at redshift $z_l$ and an observer, in order to derive the lensing properties \citep{SchneiderEhlersFalco1992}. We always assume that the lens is spherically symmetric, since ellipticity does not significantly affect the statistics of image separations \citep{Huterer2005}.

\subsubsection{Surface Density}

The surface density $\Sigma$ can be computed by integrating the 3D density profile of the halo $\rho(r)$ over the parallel coordinate along the line-of-sight, and expressed as a function of the perpendicular coordinate in the lens plane (thin lens approximation)
\begin{equation}
	\Sigma(R) = 2 \int_R^{\infty} dr \frac{r}{\sqrt{r^2 - R^2}} \rho(r) .
\label{eq:Surface_Density}
\end{equation}
The condition for strong lensing to occur is that the surface mass density exceeds the critical threshold (critical surface density)
\begin{equation}
	\Sigma_c = \frac{c^2}{4\pi G} \frac{D_s}{D_{ls}D_l},
\end{equation}
which solely depends on the angular diameter distances from the observer to the lens and source plane, corresponding to $D_l$ and $D_s$ respectively, as well as the angular distance between lens and source plane $D_{ls}$. The angular diameter distance is given by
\begin{equation}
	D_{i} = \frac{1}{(1 + z_i)}\int_0^{z_i}\frac{c dz}{H(z)}
\end{equation}
This expression holds in the case where a flat cosmology is assumed.

\begin{figure*}
\includegraphics[width=1.0\textwidth, height=0.2\textheight]{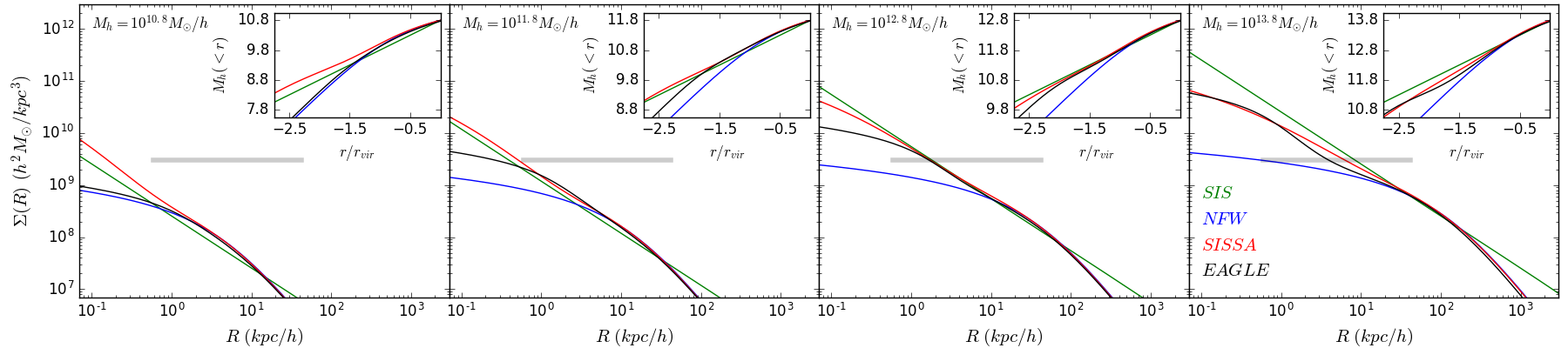}
\caption{Surface mass density as a function of the radial distance in the lens plane for the different lens models: \textbf{SIS} (green line), \textbf{NFW} (blue line), \textbf{SISSA} (red line) and a halo profile derived from the \textbf{EAGLE} simulation (black line). The grey solid line corresponds to the critical surface density $\Sigma_{c}$ for $z_l=0.5$ and $z_s = 2.0$. The figure insets show the mass enclosed within radius $r$, where the x-axis is scaled by the virial radius $r_{\textit{vir}}$.}
\label{fig:HALO_DENSITY_PROFILES}
\end{figure*}

Figure~\ref{fig:HALO_DENSITY_PROFILES} shows the radial dependence of the surface mass density for the various halo density profiles that were considered in this work. The critical surface density, when the source is at redshift $z_s = 2.0$ and the lens at $z_l = 0.5$, is also shown in the figure as the grey solid line. The different panels of the figure correspond to different halo masses (shown in their upper left corner). Note that the maximum resolution of the EAGLE simulation is $\sim 1 \,$kpc, below which there is no guarantee their fit is realistic. Each panel has an inset plot showing the mass enclosed within a certain radius.

In low mass halos $(M_h <10^{11.5} \, M_{\odot})$ the predictions from the EAGLE simulation agrees very well with the NFW profile. This range of halo masses corresponds to dwarf galaxies, where the baryon fraction of stellar to halo mass is very low and the dark matter dominates the mass budget. The critical surface density indicated that halos in this range are very inefficient lenses, not being able to produce multiple images. The SISSA model still predicts that there are baryons in these halos, but concentrated in lower radial scales beyond the probed range of the EAGLE simulation.

In intermediate mass halos $10^{11.5} \, M_{\odot}<M_h <10^{13.5} \, M_{\odot}$ the EAGLE density profile gradually departs from the NFW model as baryons start to play an important role. This range of halo masses corresponds to typical early-type galaxies, where the baryon fraction peaks causing baryonic effects to be more prominent. The dense central regions in these objects, which result from the contribution of baryons, makes them very efficient lenses. There is a fairly good agreement between the EAGLE model and both SIS and SISSA models in this range.

In high mass halos $M_h >10^{13.5} \, M_{\odot}$ the EAGLE model agrees fairly well with the NFW model for radii larger than about $\sim 10$ kpc, while their central regions are still dominated by the presence of baryons. This range of halo masses corresponds to groups/clusters of galaxies, where the baryon fraction gradually decreases until it reaches the universal mean value $f_b = \Omega_b/\Omega_m$.  The SISSA model in this range produce denser central regions as expected, since it is not intended for the description of groups/clusters of galaxies (does not account for the increase in the ratio of halo to stellar mass as the halo grows).

\subsubsection{Image Separation}

Assuming that light rays are coming from a distant point-like source and crossing the lens plane at an angular position $\theta$, they will get deflected by an angle $\alpha(\theta)$ which is given by
\begin{equation}
	\alpha(\theta|z_l,z_s,M_h) = \frac{2}{\theta}\int_0^{\theta} \theta d\theta \frac{\Sigma(D_l\,\theta|z_l,M_h)}{\Sigma_c(z_l,z_s)} .
\label{eq:Deflection_Angle}
\end{equation}
This property strongly depends on the mass enclosed within the radius $R\equiv D_l \theta$. The true and observed positions of the source in the sky are related through the simple transformation from the lens to the source plane,
\begin{equation}
	\beta(\theta)=\theta - \frac{\theta}{|\theta|}\alpha(|\theta|) \label{eq:Lens_Equation}
\end{equation}
referred to as the lens equation. The solutions of the lens equation $\theta_i$, given the position of the source $\beta$ in the source plane, will give the positions of the lensed images in the lens plane. The magnification of individual images can then be computed from
\begin{equation}
	\mu(\theta_i | z_l,z_s,M_h) =\frac{1}{\lambda_r \lambda_t}, \label{eq:magnification}
\end{equation}
with
\begin{equation}
	\lambda_{r, t} = 1 - \kappa(\theta_i) \pm \gamma(\theta_i), \label{eq:crtitical_caustic_lines}
\end{equation}
where the quantities $\kappa(\theta) = \Sigma(\theta) / \Sigma_c$ and $\gamma(\theta) = \alpha(\theta) / \theta - \kappa(\theta)$ are the convergence and shear, respectively, given as a function of the angular position in the lens plane. Therefore, the total magnification of the source, at position $\beta$ in the source plane, is computed by summing up the absolute values of the magnifications of the individual images $\mu_i$ that are formed.

\begin{figure}
\includegraphics[width=1.0\columnwidth, height=0.3\textheight]{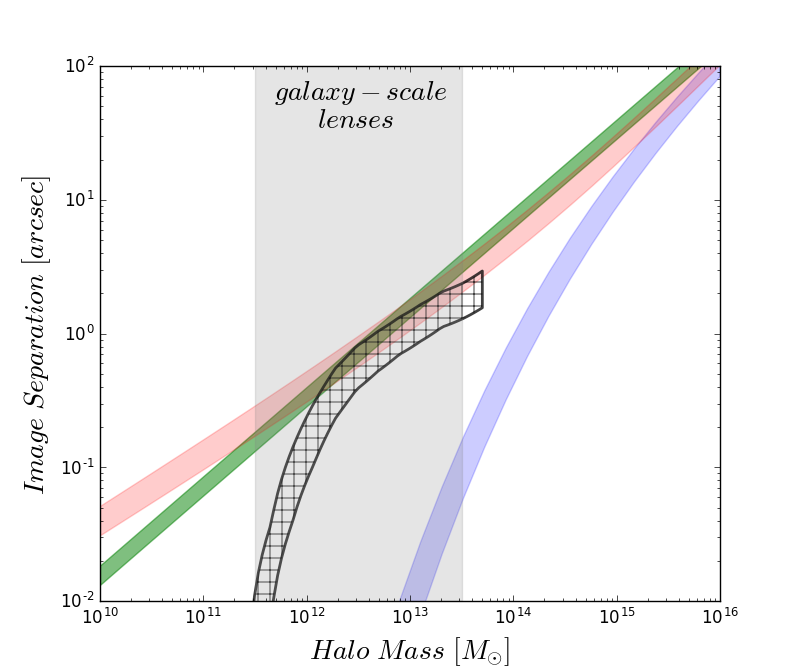}
\caption{The image separation $\theta$, as a function of the halo mass for the different lens models: \textbf{SIS} (green), \textbf{NFW} (blue), \textbf{SISSA} (red) and \textbf{EAGLE} (black hatched). The width of the stripes correspond to a lens redshift range $z_l = 0.5 - 1.0$, while the redshift of the source is kept fixed to $z_s = 2.0$. The virialization redshift is assumed to be equal to the redshift of the lens $z_{l,v}  = z_l$ in this case.}
\label{fig:IMAGE_SEPARATION}
\end{figure}
The quantities $\lambda_{r, t}$ in the denominator of Eq.~\ref{eq:magnification} define the radial and tangential critical curves in the lens plane, where the magnification diverges (when $\lambda_{r, t}$ become zero). The Einstein radius for a specific halo density profile corresponds to the radius tangential critical curve, from which we compute the image separation for a set of lens and source parameters as twice it's value. Figure~\ref{fig:IMAGE_SEPARATION} shows how the image separation changes as a function of the halo mass for the different halo models. We can see that EAGLE predicts far smaller image separations for lenses with a mass $10^{11.5} M_{\odot} < M_h < 10^{12.5} M_{\odot}$ compared to the SIS and SISSA models while in the range $10^{12.5} M_{\odot} < M_h < 10^{13.5} M_{\odot}$ there is a good agreement.

\subsubsection{Cross-Section}

\begin{figure}
\includegraphics[width=1.0\columnwidth, height=0.3\textheight]{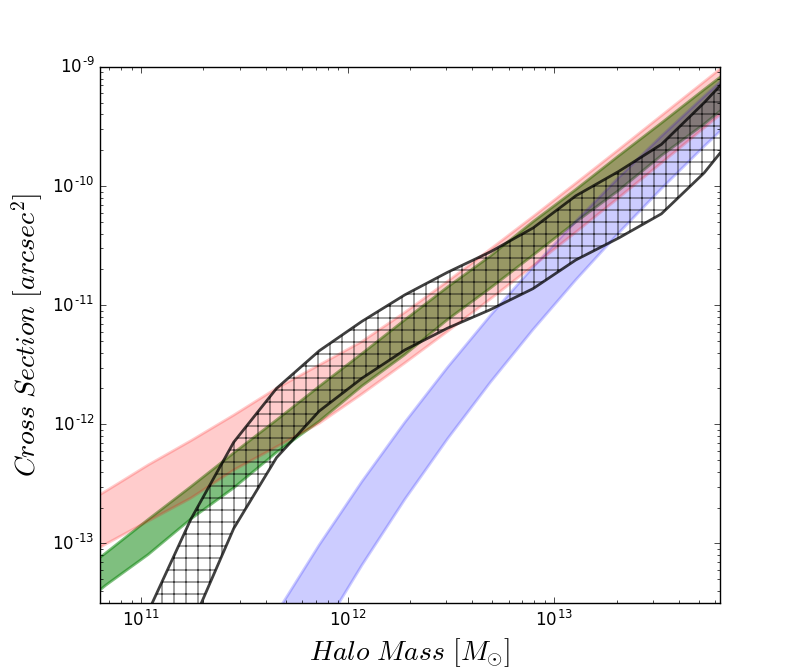}
\caption{ The cross section $\sigma(\mu>2)$, as a function of the halo mass for the different lens models: \textbf{SIS} (green), \textbf{NFW} (blue), \textbf{SISSA} (red) and \textbf{EAGLE} (black hatched). The width of the stripes correspond to a lens redshift range $z_l = 0.5 - 1.0$, while the redshift of the source is kept fixed to $z_s = 2.0$. The virialization redshift is assumed to be equal to the redshift of the lens $z_{l,v}  = z_l$ in this case. The range of halo mass corresponds to the grey highlighted area in Figure~\ref{fig:IMAGE_SEPARATION} of galaxy-scale lenses.}
\label{fig:CROSS_SECTION}
\end{figure}
The most important quantity for studies of strong lens statistics is the cross section. This is defined as the area in the source plane where the source has to lie in order to have a total magnification of $>\mu$. For a spherically symmetric mass distribution the cross section can be easily computed by
\begin{equation}
	\sigma(\geq\mu,z_l,z_s, M_h) = \pi \, \beta^2(\mu),
\end{equation}
where  $\beta(\mu)$ is the radius in the source plane corresponding to a magnification  $\mu$.

We calculated the cross section using a
minimum magnification factor of $\mu_{min} = 2$. For the SIS model,
this
corresponds to the
strong-lensing regime in which multiple images are produced. We used the same
minimum magnification factor for the other density profiles, even though this
is not the magnification at which multiple images start to be seen.
This was partly for consistency
but also because we did not originally select our sample of lensed sources because they
had multiple images but because their flux densities were amplified enough to be detected
in a sample of bright 500-$\mu$m sources.

Figure~\ref{fig:CROSS_SECTION} shows the behaviour of the cross section as a function of the halo mass for the different halo models, only for the range of galaxy-scale lenses. As illustrated in the figure, for the range of masses relevant to galaxy-scale lenses $10^{11.5} M_{\odot} < M_h < 10^{13.0} M_{\odot}$, there is an agreement between the SIS, SISSA and EAGLE models. As the halo mass grows above $10^{13.0} M_{\odot}$ the EAGLE's cross section behaviour starts to divert from these and slightly becomes similar to that of the NFW model.

\subsubsection{Magnification Bias}

 \begin{figure}
\includegraphics[width=0.95\columnwidth, height=0.25\textheight]{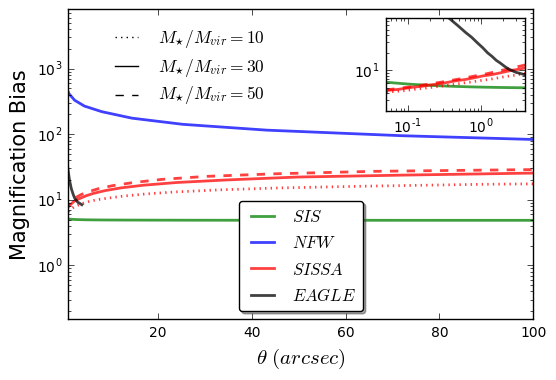}
\caption{The magnification bias as a function of the image separation, computed for a luminosity function $\Phi(L) \propto L^{-2.1}$. The calculation is performed for different lens model : \textbf{SIS} (green), \textbf{NFW} (blue), \textbf{SISSA} (red) and \textbf{EAGLE} (black). The various red lines correspond to the SISSA model adopting different choices for the ration of stellar to halo-mass. The inset plot show a zoom in to the smaller angular scales.}
\label{fig:MAGNIFICATION_BIAS}
\end{figure}

'Magnification bias' leads to lensed systems
being over-represented in a flux-limited
or magnitude-limited sample because there are more low-luminosity sources in the
universe, which lensing can boost over the flux limit, than high-luminosity sources \citep[e.g.][]{Mason2015, Eales2015}. Our study is immune to this effect because
our statistical methodology (Section 3.4) is based on the assumption that
we have found (it doesn't matter in what way) a sample of lensed sources, and
we then consider the conditional probability of the
the Einstein radius given a particular source redshift. However, because
shallower density distributions produce larger magnifications,
magnification
bias could potentially
distort the distributions of Einstein radii that we measure.
We have modelled this bias in the following way.

The magnification bias causes sources that are fainter than the
limiting magnitude of the survey to be detected in the sample. We define this bias
factor as
\begin{equation}
	B(L | z_s) = \frac{2}{\beta_r^2} \int_0^{\beta_r} \beta \frac{\Phi(L / \mu(\beta)| z_s)}{\Phi(L | z_s)} \frac{d\beta}{\mu(\beta)}
\end{equation}
where $\Phi(L | z_s)$ is the luminosity function. We calculate how the
bias factor depends on image separation for the different density distributions.
We assume that the luminosity function follows
a power-law with the form
$\Phi(L | z_s) \propto L^{-2.1}$, which is a good approximation to the
form of the submillimetre luminosity function at high luminosities
\citep{Gruppioni2013}, and we assume this is the same for all source redshifts.

Figure \ref{fig:MAGNIFICATION_BIAS} shows the computed magnification bias as a function of the image separation for the different lens models. Although, in principle, we could use our models to correct for this effect, we have decided not to do this because the luminosity function for submm sources is still very poorly constrained, and so the model is very uncertain. Figure 5 shows that there will be no effect for the SIS model, because the magnification bias is independent of angular image separation, but the effect for the other density profiles may be significant.

\subsection{Formalism of Strong Lens Statistics}  \label{sec:subsection_3.4}

We adopt the standard formalism for computing lensing statistics \citep{TurnerOstrikerGott1984}, where we consider a population of dark-matter halos that act as deflectors located at redshift $z_l$ and can be characterised by their mass $M_h$. The differential probability that a source at redshift $z_s$ is strongly lensed with total magnification $\geq \mu$ by that population of deflectors is given by
\begin{equation}
	\frac{dP}{dz_ldM_h} = \frac{d^2N}{dM_hdV} \frac{d^2V}{dz_ld\Omega} \sigma(\geq\mu, z_l, z_s, M_h),
\label{eq:differential_probability_distribution}
\end{equation}
where
\begin{equation}
	 \frac{d^2V}{dz_ld\Omega} = \frac{c}{H_0} \frac{(1+z_l)^2 D_A^2(z_l)}{E(z_l)}
\label{eq:}
\end{equation}
is the comoving volume element per unit of $z_l$-interval and solid angle, while $d^2N / dM_hdV$ is the number density of deflectors per units of $M_h$-interval at different redshifts

The total lensing probability $P(z_s, \geq\mu)$ can be computed by integrating Eq.~\ref{eq:differential_probability_distribution} over the lens redshift and halo mass ranges. To calculate the probability distribution of image separations we insert a selection function in the integral in order to select only the combination of parameters that produce image separations in the interval $\theta \pm d\theta$. The probability distribution as a function of the image separation then becomes
\begin{equation}
	P(\theta \, | \, z_s, \geq\mu) = \int_0^{z_s} dz_l \int_{0}^{\infty} dM_h \frac{dP}{dz_ldM_h} \delta[\theta - \tilde{\theta}(z_l, z_s, M_h)],
\label{eq:probability_distribution}
\end{equation}
where $\tilde{\theta}(z_l, z_s, M_h)$ is calculated for each model as twice the Einstein radius (tangential critical curve) and the Dirac $\delta$-function is unity if the combination of parameters corresponds to image separation $\tilde{\theta}$ in the interval  ($\theta - d\theta$,  $\theta + d\theta$).

The amplitude of the image separation distribution in Eq.~\ref{eq:probability_distribution} increases with increasing source redshift independently of the angular scale, since we sample a larger volume of the universe. The normalised image separation distribution on the other hand,
\begin{equation}
	p(\theta \, | \, z_s, \geq\mu) = \frac{ P(\theta \, | \, z_s, \geq\mu) }{ \int_0^{\infty} d\theta  P(\theta \, | \, z_s, \geq\mu) },
\label{eq:normalized_probability_distribution:}
\end{equation}
is quite insensitive to the source population as well as the cosmological parameters \citep{Oguri2002b}. Comparing the predicted normalised distribution with the observed one, we therefore probe the combination of the halo mass function and density profiles of halos which affect the shape of the distribution.

In our analysis we assume a two-transitiotn mass model, following the methodology adopted in previous studies \citep{PorcianiMadau2000, KochanekWhite2001, Oguri2002b, Kuhlen2004}. This approach was introduced in order to account for baryons, which probably affect the shape of halo's density profile by means of adiabatic contraction \citep{Blumenthal1986} and cooling \citep{WhiteRees1978} when the baryon fraction is relatively high. In our model, halos below the mass $M_{\textit{min}}$ (corresponding to dwarf galaxies) and above $M_{\textit{max}}$ (corresponding to clusters of galaxies), are described by the NFW profile to account for the expected low baryon fraction.In the intermediate mass range (corresponding to early-type galaxies) halos are described by either the SIS or SISSA model, where the baryon fraction is expected to reach the peak.

Another quantity that was introduced in the analytic description of the SIS and SISSA models in Lapi et al. (2012), is the virialization redshift of the lens $z_{l,v}$. According to their study, the frequently made approximation $z_{l,v} \approx z_l$ leads to an underestimate of the lensing probability. This is because a lower value of the virialization redshift leads to an overestimation of the halo size and therefore to an underestimation of the halo's density. As a result, a higher upper-transition mass would be necessary in order to match the observed distribution of image separations. We examine the effect of the virialization redshift on the transition-masses of our model by considering both a $z_{l,v} = z_l$ and $z_{l,v} = 2.5$ (see Lapi et al. 2012 for details) when computing the theoretical distribution of image separations.


\section{Results}  \label{sec:section_4}

In this section we follow the methodology described in Section~\ref{sec:section_3}, to derive the theoretical distributions of image separations. We then compare our model predictions with the normalised histogram of the observed image separations for two samples of Herschel selected lensed sources. We emphasise that the use of the conditional probability distribution means that our analysis is independent of the properties of the source population. We carry out the analysis separately for the sample of sources observed with ALMA and SMA.

\subsection{Comparison with Observations}  \label{sec:subsection_4.2}

\begin{figure*}
\includegraphics[width=1.0\textwidth, height=0.35\textheight]{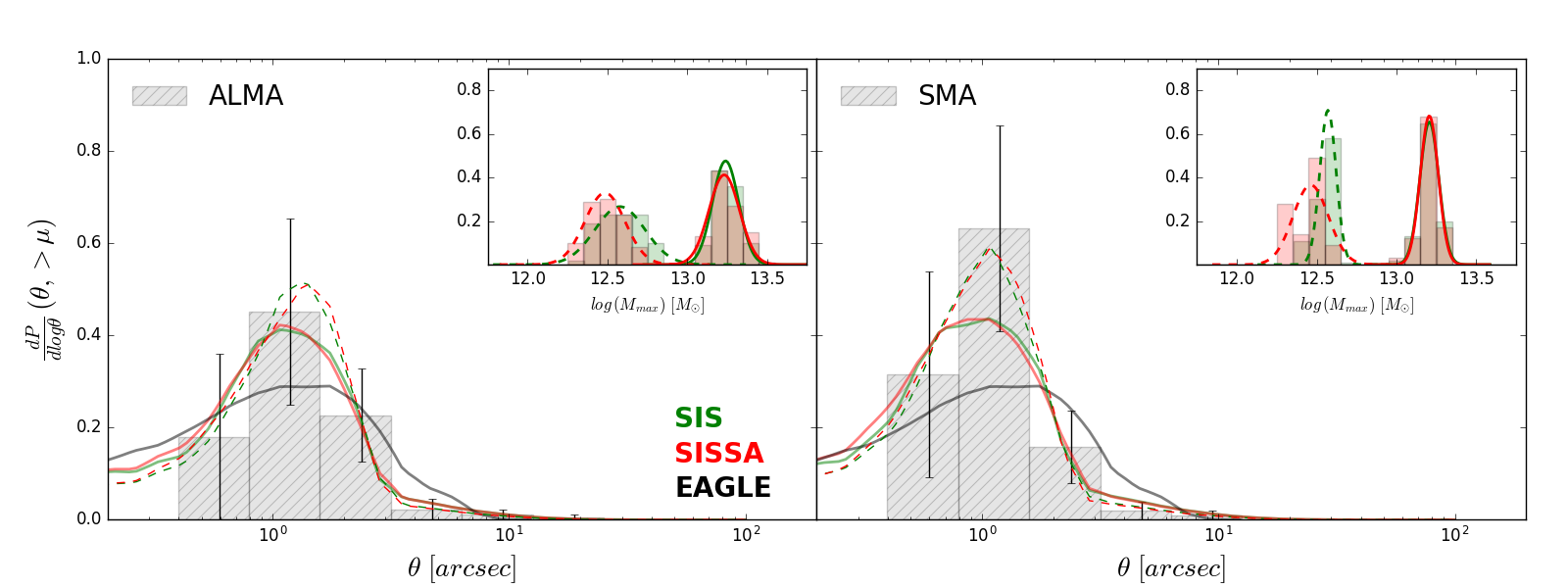}
\caption{The predicted distribution of image separations adopting either the SIS (green) or SISSA (red) profiles for galaxy-scale lenses and following the procedure described in Section~\ref{sec:subsection_4.2}. The predicted distribution of image separation, which was derived assuming a halo model calibrated from the EAGLE simulation results, is shown with black dashed lines. Left and right panels correspond to the fits with the two samples of lenses followed-up with ALMA and SMA, respectively. The gray-scale histograms are the observed distributions of our samples. The figure insets show the distribution of the upper mass-transition parameter after performing $\sim$100 realisations. The predictions adopting a virialization redshift $z_{l,v} = z_l$ are shown as straight lines while the ones with a virialization redshift $z_{l,v} = 2.5$ are shown as dashed lines.}
\label{fig:RESULTS}
\end{figure*}

\begin{table*}
	\centering
	\caption{ Best-fit value of the two transition masses that were used in our analytic model, adopting either the SIS or SISSA model for the description of galaxy-scale lenses. These values were derived assuming a virialization redshift $z_{l,v} = z_l$ for the first two rows and $z_{l,v} = 2.5$ for the last two. }
	\begin{tabular}{lccccccc}
	\hline
	\hline
		  & $\quad\quad$ & log$(M_{\textit{min}})_{\rm SIS}$ & $\quad\quad$ & log$(M_{\textit{max}})_{\rm SIS}$ $\quad\quad$ & $\quad\quad$ log$(M_{\textit{min}})_{\rm SISSA}$ & $\quad\quad$ & log$(M_{\textit{max}})_{\rm SISSA}$ \\
	\hline
	ALMA$_{z_{vir} = z_l}$ & $\quad\quad$ & $\le$12.4 & $\quad\quad$ & 13.25 $\pm$ 0.10 $\quad\quad$ & $\le$12.3 & $\quad\quad$ &13.20 $\pm$ 0.11 \\
	SMA $_{z_{vir} = z_l}$  & $\quad\quad$ & $\le$12.2 & $\quad\quad$ & 13.19 $\pm$ 0.07 $\quad\quad$ & $\le$12.0 & $\quad\quad$ &13.20 $\pm$ 0.06 \\

	ALMA$_{z_{vir} = 2.5}$ & $\quad\quad$ & $\le$12.1 & $\quad\quad$ & 12.56 $\pm$ 0.13 $\quad\quad$ & $\le$12.1 & $\quad\quad$ & 12.48 $\pm$ 0.10 \\
	SMA$_{z_{vir} = 2.5}$   & $\quad\quad$ & $\le$11.9 & $\quad\quad$ & 12.54 $\pm$ 0.07 $\quad\quad$ & $\le$11.9 & $\quad\quad$ & 12.42 $\pm$ 0.10 \\
	\hline
	\hline
	\end{tabular}
	\label{tab:best_fit_parameters}
\end{table*}

We derive the values of our two transition-mass models, described in Section~\ref{sec:subsection_3.4}, by performing a standard $\chi^2$ minimisation method
\begin{equation}
	\chi^2 = \sum_{i}\frac{( P_{}(\theta_i | \geq\mu)  - P^{'}_{}(\theta_i | \geq\mu))^2 }{\sigma_{}^2(\theta_i)},
\end{equation}
where $P_{}(\theta)$ and $P^{'}_{}(\theta)$ are the observed and theoretical normalised image separation distributions respectively. The quantity $\sigma_{}(\theta)$ is the standard deviation of each bin of the observed histogram of image separations, which is derived from poisson statistics.

Figure~\ref{fig:RESULTS} show a comparisons of the observed and predicted distributions of image separations. The black solid line shows the predicted distribution using the analytic mass density distribution obtained from the EAGLE simulation (Eq.~\ref{eq:EAGLE}). This agrees fairly well with the observations, and does not require the imposition of transition masses. The other lines show the predictions of our analytic models with two transition masses. The graphs
show our predictions adopting a virialization redshift $z_{l,v} = z_l$ and $z_{l,v} = 2.5$ as straight and dashed lines, respectively.

The grey histograms in each graph correspond to the observed distributions for the sample of sources observed with ALMA on the left-hand side and with the sample of sources observed with SMA that was used in our previous study (Eales 2015), on the right-hand side. The best-fit values of the two transition-masses are shown in Tables \ref{tab:best_fit_parameters} for the two different choice of virialization redshift along with the different choices of halo density profiles and observed sample. In order to account for the uncertainty on the measured image separations, we perform 100 simulations for each measurement by resampling each value at random from a Gaussian distribution with a standard deviation equal the value's error. For each realisation of the observed distribution we perform the above fitting procedure and we end up with a distribution for the upper transition-mass from which we derive it's errors.

In our analysis we decided to exclude the object J141351.9-000026, which as discussed in Section~\ref{sec:section_2} has a very large Einstein radius as a result of lensing by a galaxy cluster. If we were to include this object in the analysis there wouldn't be any significant difference in the constrained value of the Maximum transition masses. This is because the constrain is more sensitive to the contribution from the galaxy scale lenses. Increasing the Maximum transition mass will shift the kink of the distribution to larger scales and the lack of objects in that range constrains it's value. Including an object with significantly larger Einstein radius than where the kink is observed will not significantly contribute to the fitting method. Furthermore, no proper modelling has been performed for this object to extract the value of it's Einstein radius.

Predictions adopting either of the analytic profiles, SIS and SISSA, as well as the density profile derived from the EAGLE simulation, seem to be in good agreement with observations. Furthermore, comparing the fitted values of the upper transition mass that were obtained for the different samples of lenses, we find a slight difference that is not significant (i.e. < $1\,\sigma$). As mentioned in Section~\ref{sec:section_2}, the observed distribution of image separations, for the SMA sample, is biased towards lower angular separations, which leads to an underestimate of the upper transition-mass. Concerning the lower transition-mass, we are still not in a position to set good constrains because our fitting method cannot distinguish models with $M_{min} \lesssim 10^{12.5} M_{\odot}$. Finally, the virialization redshift strongly affects the resulting transition masses, pushing them to lower values. However, there is still no evidence to support such a low-transition mass between galaxies and clusters.


\section{Discussion \& Conclusions}  \label{sec:section_5}

Wide-area extragalactic surveys conducted at submillimeter wavelengths has allowed us to discover a new population of strongly lensed galaxies \citep{Negrello2010, Negrello2017, Nayyeri2016}. Their potential to produce very large samples of strong lenses \citep{Gonzalez-Nuevo2012} and the simplicity of the selection function \citep{Blain1996, Perrotta2002, Perrotta2003, Negrello2007}, will greatly benefit the study of strong lens statistics, a subject which has previously been studied by optical \citep{Bolton2006, More2012} and radio surveys \citep{Browne2003, Oguri2006a}.
Extragalactic surveys undertaken with Herschel Space Observatory have demonstrated the potential of this method by producing large samples of candidate strong lenses \citep{Wardlow2013, Nayyeri2016, Negrello2017}. We carried out follow-up observations with ALMA of 16 candidate strongly lensed Herschel sources, selected from the H-ATLAS and HeLMS surveys, expecting that based on their bright 500-$\mu m$ flux densities that they should be lensed. Out of these sources, 15 show clear evidence of lensing features.

In this study we predict the distribution of image separations of strongly lensed systems produced by a population of dark matter halos parametrised by the halo mass function derived from hydrodynamical cosmological simulations \citep{Bocquet2016}. The largest uncertainty that enters the calculation of the theoretical image separation distribution is the total mass distribution of these halos, which is the primary focus of this study. For the first time we used a halo density profile that was derived from the EAGLE simulation \citep{Schaller2015a, Schaller2015b}, which is calibrated so that it provides a good fit across a wide range of halo masses. We showed that the combination of mass density distributions and the halo mass function predicted by cosmological numerical simulations can reproduce the observed distribution of image separation of strong lenses found in submillimeter surveys.

We also consider a different approach adopting analytical recipes for the description of the total mass distribution in dark-matter halos. Since there is not a single analytic model to describe halo density profiles across the whole range of halo masses we introduce two transition masses between dwarf to early-type galaxies and early-type to cluster of galaxies, respectively. For the description of early-type galaxy halos we consider two approaches, the SIS and SISSA models, while for dwarfs and cluster of galaxies we adopt the NFW model. We utilise our samples of strong lenses from which we derive the observed distribution of image separation, in order to constrain the values of the transition masses. We were able to set good constrains on the maximum transition-mass (see Table~\ref{tab:best_fit_parameters}). Our results agree with previous studies of strong lens statistics using the CLASS \citep{Myers2003,Browne2003} sample of strong lenses, where they place the value of the upper transition-mass at $\sim 10^{13}M_{\odot}$ \citep{PorcianiMadau2000, KochanekWhite2001, Oguri2002b, LiOstriker2002, Kuhlen2004}. A complementary approach was adopted by \cite{Oguri2006b} in which the author used a two-component halo density profile, comprised of an NFW dark matter halo and a Hernquist model for the central galaxy, that also considers the effect of adiabatic contraction of dark matter. This profile has a smooth transition between galaxy and cluster scale lenses and does not require the assumption of a transition mass and has the potential to better account for the contribution from group-scale lenses. This profile seem to provide a relatively good fit to radio \citep{Oguri2006b} and optical data \citep{More2012}. However, as our sample is still limited in numbers to make such distinctions between models, we have not considered this approach.

A larger sample is also required in order to distinguish between models with a minimum transition mass $< 10^{12} M_{\odot}$ \citep{Ma2003}. However, our candidate sample selection does not have any completeness issues at low angular resolutions as optical surveys do \citep{More2016}. This is because our selection is purely flux based and does not require the identification of individual multiply lensed images. Since our sample has no biases at small angular separation, follow up observations with ALMA can in fact probe the subarcsec scale of the image separation distribution (see e.g HeLMS J235331.9+031718).

We also examined the effect of varying the virialization redshift of the lens $z_{l,v}$, which is one of the parameters of our analytic models. Previous studies of strong lens statistics have ignored it's effect and always assumed that it coincides with the actual redshift of the halo $z_{l,v} = z_l$. Lapi et al. 2012 argue that this approximation leads to an overestimate of the halo's size and, subsequently, to an underestimate of the lensing probability. We showed that adopting the value suggested by Lapi et al. 2012, $z_{l,v} = 2.5$, the constrained value of the maximum transition mass significantly decreases (see Table~\ref{tab:best_fit_parameters}).

This approach of predicting the distribution of image separation based on the population of dark-matter halos selected on the basis of their halo mass, provides a confirmation of the standard cold dark-matter paradigm. However, the current samples of strong lenses are still not large enough in order to able to distinguish between the different models that attempt to describe the internal structure of these halos. Scaling from the errors in Figure~\ref{fig:RESULTS} we estimate that a sample of $\sim 500$ would be required for this distinction to be made possible.

Is it practical to produce such a large sample of lensed sources.
Gonzalez-Nuevo et al. (2012) have proposed a method for finding at least 1000 lensed sources from
the Herschel surveys. However, their method is based on finding galaxies that lie
close to the position of a Herschel source, and therefore
have a high probability of being associated with it, but which
have much lower estimated redshifts than
the Herschel source. This method will therefore be biased towards lensing systems with
small image separations and so is
not suitable for our purpose.

The most promising method is a variant of the method used by Negrello et al. (2010). There are only $\simeq$150 probable lensed sources with the 500-$\mu$m flux densities $>$100 mJy (Nayyeri et al. 2016; Negrello et al. 2017), the cutoff used by Negrello et al. (2010). However, Negrello et al. (2010) estimate that the fraction of high-redshift Herschel sources that are strongly lensed is $>$50\% down to a 500-$\mu$m flux density of $\simeq$50 mJy. We have shown  in this paper that observations with ALMA with exposure times of only a few minutes are enough to show that a bright Herschel source is lensed. Therefore, a programme to obtain short ALMA continuum observations of 500-1000 bright Herschel sources seems a practical way of assembling the required sample of 500 lensed systems. The more challenging part of the programme would be to obtain redshifts for the sources.  However,15-minute ALMA observations are often enough to obtain a redshift for a bright Herschel source. Therefore, even this part of the project seems practical in an ALMA Large Programme. In the slightly longer term, continuum surveys with the Square Kilometre Array will contain tens of thousands of lensed sources (Mancuso et al. 2015).

\section*{Acknowledgements}
MN acknowledges financial support from the European Union's Horizon 2020 research and innovation programme under the Marie Sk{\l}odowska-Curie grant agreement No 707601.
Some of the spectroscopic redshift reported in this paper were obtained with the Southern African Large Telescope (SALT) under proposal 2015-2-MLT-006, PI: Stephen Serjeant.
LM acknowledged support from the South African Department of Science and Technology and the SA National Research Foundation.
EV acknowledges funding from STFC consolidated grant ST/K000926/1. MS and SAE have received funding from the European Union Seventh Framework Programme ([FP7/2007-2013] [FP\&/2007-2011]) under grant agreement No. 607254.
SJM, LD and PJC acknowledge support from the European Research Council (ERC) in the form of the Consolidator Grant COSMICDUST (ERC-2014-CoG-647939, PI H.L.Gomez).
SJM, LD and RJI acknowledge support from the ERC in the form of the Advanced Investigator Programme, COSMICISM (ERC-2012-ADG\_20120216, PI R.J.Ivison).




\appendix

\section{Understanding the SISSA Model} \label{sec:Appendix_A}

In this section we show the effects of the various ingredients that enter the calculation of cumulative image separation distribution. This is calculated from Eq.~\ref{eq:probability_distribution} by substituting the Dirac delta function by the Heaviside step function. For this particular calculation only we use the standard method for computing cross-sections as $\sigma = \pi \beta_{\textit{cr}}^2$, where $\beta_{\textit{cr}}$ is the radial caustic within which multiple images are formed.

\begin{figure*}
\centering
\textbf{\Large$\textit{SISSA + NFW}$}\par\medskip
\includegraphics[width=0.85\textwidth, height=0.35\textheight]{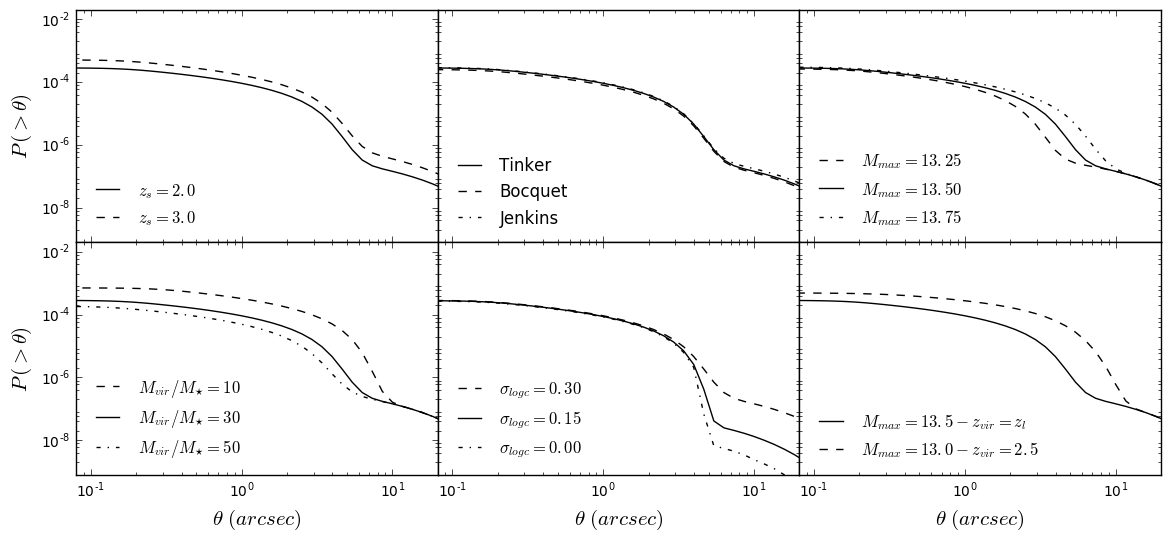}
\caption{ Effects of parameter variation in the cumulative distribution of image separations.}
\label{fig:IMAGE_SEPARATION_CUMULATIVE_DISTRIBUTION_SISSA}
\end{figure*}

\subsection{Variation in $z_s$}
The source redshift, $z_s$, predominantly affects the amplitude of the distribution. This is to be expected since a higher source redshift corresponds to a larger volume of the Universe being considered. However, the predicted distributions of image separations in Section~\ref{sec:section_4} are normalised and therefore this additional factor cancels out. 

\subsection{Variation in Halo-Mass Function}

The use of different halo mass functions models has very little effect on the distribution of image separations. The T08 and Bocquet mass functions assume the same formalism but their parameters are calibrated from DM only and Hydro cosmological simulations, respectively. Comparing the halo mass functions themselves we find that the effect of baryons is to suppress only slightly the creation of massive halos but only as small redshifts. At higher redshifts they tend to agree fairly well.

\subsection{Variation in $M_{\textit{max}}$}
The upper transition mass $M_{\textit{max}}$ parametrizes the change from galaxy-sized SISSA to group- and cluster-sized NFW lenses. This parameter determined the position of the kink in the image separation distribution. For an upper transition mass of $logM_{\textit{max}} = 13.50$ this transition occurs at $\theta = 7''$. Lowering the transition mass to $logM_{\textit{max}} = 13.25$ shifts this transition down to $\theta = 4''$, while increasing it to $logM_{\textit{max}} = 13.75$ this transition shifts up to $\theta = 10''$.

\subsection{Variation in $M_{\textit{vir}}/M_{\star}$ Ratio}
The ratio between the halo and stellar mass $M_{\textit{vir}}/M_{\star}$, is an important parameter in the SISSA model and it's effect on the distribution of image separations is twofold. First, we see that increasing this ratio from 10 to 50, the abundance of arcsec-scale lenses decreases by almost a factor of $\sim 5$. Secondly, it affects the kink of the distribution by shifting it from $\theta = 5''$ to $\theta = 10''$.

\subsection{Variation in $\sigma_{\textit{logc}}$}
The parameter $\sigma_{\textit{logc}}$ controls the standard deviation of the distribution of concentration parameters. This distribution is expected to have a scatter that is well described by a lognormal distribution,
\begin{equation}
p(c) = \frac{1}{\sqrt{2 \pi} \sigma_{\textit{logc}} c} exp\left[ - \frac{\left( logc - log\bar{c} \right)^2}{2 \sigma_{\textit{logc}}^2} \right] \, ,
\end{equation}
where the $\bar{c}$ is given by Eq.~\ref{eq:concentration_Prada2011}. The SIS model does not depend on this parameter and therefore arcsec-scale lenses produced by galaxies adopting this model, are not affected by any changes \citep{TakahashiChiba2001, Kuhlen2004, Oguri2006b}. However this parameter does enter in the SISSA model through the NFW component. Although, it's effect is not as drastic as it is for the wide-separation lenses produce by galaxy groups and cluster adopting a pure NFW model, it's still affects the resulting distribution of image separation by shifting the kink by a few arcsec.

\subsection{Variation in $z_{\textit{vir}}$}

As described in Section~\ref{sec:subsection_3.1} the commonly made approximation that the virialization redshift in equal to the observed redshift lead to an overestimation of the halo size and therefore a decrease of the halo's density, making halos less efficient. Adopting a virialization redshift $z_{l,v} = 2.5$ drastically shifts the kink of the distribution to larger angular scales as well as it increases the abundance of galaxy-scale lenses. In this case the virialization redshift is introduced only for the SISSA model, as it would be unrealistic to assume that group- and cluster-scale lenses had beed virialiazed at such high redshift.

\bsp	
\label{lastpage}

\begin{thebibliography}{99}

\bibitem[\protect\citeauthoryear{Blain}{1996}]{Blain1996}
{Blain}, A.~W., \href{http://adsabs.harvard.edu/abs/1996MNRAS.283.1340B}{1996}, MNRAS, 283, 1340

\bibitem[\protect\citeauthoryear{Blumenthal}{1986}]{Blumenthal1986}
{Blumenthal}, G.~R., {Faber}, S.~M., {Flores}, R., {Primack}, J.~R., \href{http://adsabs.harvard.edu/abs/1986ApJ...301...27B}{1986}, ApJ, 301, 27

\bibitem[\protect\citeauthoryear{Bocquet et al.}{2016}]{Bocquet2016}
{Bocquet}, S., {Saro}, A., {Dolag}, K., {Mohr}, J.~J., \href{http://adsabs.harvard.edu/abs/2016MNRAS.456.2361B}{2016}, MNRAS, 456, 2361

\bibitem[\protect\citeauthoryear{Bolton et al.}{2006}]{Bolton2006}
{Bolton}, A.~S., {Burles}, S., {Koopmans}, L.~V.~E., {Treu}, T., {Moustakas}, L.~A., \href{http://adsabs.harvard.edu/abs/2006ApJ...638..703B}{2006}, ApJ, 638, 703

\bibitem[\protect\citeauthoryear{Brownstein et al.}{2012}]{Brownstein2012}
{Brownstein}, J.~R., {Bolton}, A.~S., {Schlegel}, D.~J., {Eisenstein}, D.~J., {Kochanek}, C.~S., et al. \href{http://adsabs.harvard.edu/abs/2012ApJ...744...41B}{2012}, ApJ, 744, 41

\bibitem[\protect\citeauthoryear{Browne et al.}{2003}]{Browne2003}
{Browne}, I.~W.~A., {Wilkinson}, P.~N., {Jackson}, N.~J.~F., {Myers}, S.~T., {Fassnacht}, C.~D., et al. \href{http://adsabs.harvard.edu/abs/2003MNRAS.341...13B}{2003}, ApJ, 341, 13

\bibitem[\protect\citeauthoryear{Bryan \& Norman}{1998}]{BryanNorman1998}
{Bryan}, G.~L., {Norman}, M.~L., \href{http://adsabs.harvard.edu/abs/1998ApJ...495...80B}{1998}, ApJ, 495, 80

\bibitem[\protect\citeauthoryear{Bullock et al.}{2001}]{Bullock2001}
{Bullock}, J.~S., {Kolatt}, T.~S., {Sigad}, Y., {Somerville}, R.~S., {Kravtsov}, A.~V., et al., \href{http://adsabs.harvard.edu/abs/2001MNRAS.321..559B}{2001}, MNRAS, 321, 559

\bibitem[\protect\citeauthoryear{Bussmann et al.}{2013}]{Bussmann2013}
{Bussmann}, R.~S., {P{\'e}rez-Fournon}, I., {Amber}, S., {Calanog}, J., {Gurwell}, M.~A., et al. \href{http://adsabs.harvard.edu/abs/2013ApJ...779...25B}{2013}, ApJ, 779, 25

\bibitem[\protect\citeauthoryear{Calanog et al.}{2014}]{Calanog2014}
{Calanog}, J.~A., {Fu}, H., {Cooray}, A., {Wardlow}, J., {Ma}, B., {Amber}, S., {Baker}, A.~J., et al. \href{http://adsabs.harvard.edu/abs/2014ApJ...797..138C}{2014}, ApJ, 797, 138

\bibitem[\protect\citeauthoryear{Chae}{2003}]{Chae2003}
{Chae}, K.-H., \href{http://adsabs.harvard.edu/abs/2003MNRAS.346..746C}{2003} MNRAS, 346, 746

\bibitem[\protect\citeauthoryear{Collett}{2015}]{Collett2015}
{Collett}, T.~E. \href{http://adsabs.harvard.edu/abs/2015ApJ...811...20C}{2015}, ApJ, 811, 20

\bibitem[\protect\citeauthoryear{Cox et al.}{2011}]{Cox2011}
{Cox}, P., {Krips}, M., {Neri}, R., {Omont}, A., {G{\"u}sten}, R., {Menten}, K.~M., {Wyrowski}, F., et. al. \href{http://adsabs.harvard.edu/abs/2011ApJ...740...63C}{2011}, ApJ, 740, 63

\bibitem[\protect\citeauthoryear{Dye et al.}{2014}]{Dye2014}
{Dye}, S., {Negrello}, M., {Hopwood}, R., {Nightingale}, J.~W., {Bussmann}, R.~S., {Amber}, S., et. al. \href{http://adsabs.harvard.edu/abs/2014MNRAS.440.2013D}{2014}, MNRAS, 440, 2013

\bibitem[\protect\citeauthoryear{Eales et al.}{2010}]{Eales2010}
{Eales}, S., {Dunne}, L., {Clements}, D., {Cooray}, A., {De Zotti}, G., {Dye}, S., {Ivison}, R., {Jarvis}, M., et al. \href{http://adsabs.harvard.edu/abs/2010PASP..122..499E}{2010}, PASP, 122, 499

\bibitem[\protect\citeauthoryear{Eales}{2015}]{Eales2015}
{Eales}, S.~A., \href{http://adsabs.harvard.edu/abs/2015MNRAS.446.3224E}{2015}, MNRAS, 446, 3224

\bibitem[\protect\citeauthoryear{George et al.}{2013}]{George2013}
{George}, R.~D., {Ivison}, R.~J., {Hopwood}, R., {Riechers}, D.~A., {Bussmann}, R.~S., {Cox}, P., {Dye}, S., et al. \href{http://adsabs.harvard.edu/abs/2013MNRAS.436L..99G}{2013}, MNRAS, 436, 99

\bibitem[\protect\citeauthoryear{Gladders \& Yee}{2005}]{GladdersYee2005}
{Gladders}, M.~D., {Yee}, H.~K.~C., \href{http://adsabs.harvard.edu/abs/2005ApJS..157....1G}{2005}, ApJS, 157, 1

\bibitem[\protect\citeauthoryear{Gonzalez-Nuevo et al.}{2012}]{Gonzalez-Nuevo2012}
{Gonz{\'a}lez-Nuevo}, J., {Lapi}, A., {Fleuren}, S., {Bressan}, S., {Danese}, L., {De Zotti}, G., {Negrello}, M., et al., \href{http://adsabs.harvard.edu/abs/2012ApJ...749...65G}{2012}, ApJ, 749, 65

\bibitem[\protect\citeauthoryear{Gruppioni et al.}{2013}]{Gruppioni2013}
{Gruppioni}, C., {Pozzi}, F., {Rodighiero}, G., {Delvecchio}, I., {Berta}, S., {Pozzetti}, L., {Zamorani}, G., et al., \href{http://adsabs.harvard.edu/abs/2013MNRAS.432...23G}{2013}, MNRAS, 432, 23

\bibitem[\protect\citeauthoryear{Harris et al.}{2012}]{Harris2012}
{Harris}, A.~I., {Baker}, A.~J., {Frayer}, D.~T., {Smail}, I., {Swinbank}, A.~M., {Riechers}, D.~A., et al., \href{http://adsabs.harvard.edu/abs/2012ApJ...752..152H}{2012},

\bibitem[\protect\citeauthoryear{Huterer et al.}{2005}]{Huterer2005}
{Huterer}, D., {Keeton}, C.~R., {Ma}, C.-P., \href{http://adsabs.harvard.edu/abs/2005ApJ...624...34H}{2005}, ApJ, 624, 34

\bibitem[\protect\citeauthoryear{Inada et al.}{2012}]{Inada2012}
{Inada}, N., {Oguri}, M., {Shin}, M.-S., {Kayo}, I., {Strauss}, M.~A., et al. \href{http://adsabs.harvard.edu/abs/2012AJ....143..119I}{2012}, AJ, 143, 119

\bibitem[\protect\citeauthoryear{Kochanek \& White}{2001}]{KochanekWhite2001}
{Kochanek}, C.~S., {White}, M., \href{http://adsabs.harvard.edu/abs/2001ApJ...559..531K}{2001}, ApJ 559, 531

\bibitem[\protect\citeauthoryear{Jenkins et al.}{2001}]{Jenkins2001}
{Jenkins}, A., {Frenk}, C.~S., {White}, S.~D.~M., {Colberg}, J.~M., {Cole}, S., et al. \href{http://adsabs.harvard.edu/abs/2001MNRAS.321..372J}{2001}, MNRAS, 321, 372

\bibitem[\protect\citeauthoryear{Koopmans et al.}{2006}]{Koopmans2006}
{Koopmans}, L.~V.~E., {Treu}, T., {Bolton}, A.~S., {Burles}, S., {Moustakas}, L.~A., \href{http://adsabs.harvard.edu/abs/2006ApJ...649..599K}{2006}, ApJ, 649, 599

\bibitem[\protect\citeauthoryear{Koopmans et al.}{2009}]{Koopmans2009}
{Koopmans}, L.~V.~E., {Bolton}, A., {Treu}, T., {Czoske}, O., {Auger}, M.~W., et al., \href{http://adsabs.harvard.edu/abs/2009ApJ...703L..51K}{2009}, ApJ, 703, 51

\bibitem[\protect\citeauthoryear{Kuhlen et al.}{2004}]{Kuhlen2004}
{Kuhlen}, M., {Keeton}, C.~R., {Madau}, P., \href{http://adsabs.harvard.edu/abs/2004ApJ...601..104K}{2004}, ApJ, 601, 104

\bibitem[\protect\citeauthoryear{Lapi et al.}{2012}]{Lapi2012}
{Lapi}, A., {Negrello}, M., {Gonz{\'a}lez-Nuevo}, J., {Cai}, Z.-Y., et al., \href{http://adsabs.harvard.edu/abs/2012ApJ...755...46L}{2012}, ApJ, 192, 18

\bibitem[\protect\citeauthoryear{Li \& Ostriker}{2002}]{LiOstriker2002}
{Li}, L.-X., {Ostriker}, J.~P., \href{http://adsabs.harvard.edu/abs/2002ApJ...566..652L}{2002},ApJ, 566, 652

\bibitem[\protect\citeauthoryear{Li \& Ostriker}{2003}]{LiOstriker2003}
{Li}, L.-X., {Ostriker}, J.~P., \href{http://adsabs.harvard.edu/abs/2003ApJ...595..603L}{2003},  ApJ, 595, 603

\bibitem[\protect\citeauthoryear{Ma}{2003}]{Ma2003}
{Ma}, C.-P., \href{http://adsabs.harvard.edu/abs/2003ApJ...584L...1M}{2003}, ApJ, 584, L1

\bibitem[\protect\citeauthoryear{Mancuso}{2015}]{Mancuso2015}
{Mancuso}, C., {Lapi}, A., {Cai}, Z.-Y., {Negrello}, M., {De Zotti}, G., et al. \href{http://adsabs.harvard.edu/abs/2015ApJ...810...72M}{2015}, ApJ, 810, 72

\bibitem[\protect\citeauthoryear{Mason et al.}{2015}]{Mason2015}
{Mason}, C.~A., {Treu}, T., {Schmidt}, K.~B., {Collett}, T.~E., et al. \href{http://adsabs.harvard.edu/abs/2015ApJ...805...79M}{2015}, ApJ, 805, 79

\bibitem[\protect\citeauthoryear{Messias et al.}{2014}]{Messias2014}
{Messias}, H., {Dye}, S., {Nagar}, N., {Orellana}, G., {Bussmann}, R.~S., {Calanog}, J., et al. \href{http://adsabs.harvard.edu/abs/2014A\%26A...568A..92M}{2014}, A\&A, 568, 92

\bibitem[\protect\citeauthoryear{More et al.}{2012}]{More2012}
{More}, A., {Cabanac}, R., {More}, S., {Alard}, C., {Limousin}, M., et al. \href{http://adsabs.harvard.edu/abs/2012ApJ...749...38M}{2012}, ApJ, 749, 38

\bibitem[\protect\citeauthoryear{More et al.}{2016}]{More2016}
{More}, A., {Verma}, A., {Marshall}, P.~J., {More}, S., {Baeten}, E., {Wilcox}, J., et. al. \href{http://adsabs.harvard.edu/abs/2016MNRAS.455.1191M}{2016}, MNRAS, 455, 1191

\bibitem[\protect\citeauthoryear{Myers et al.}{2003}]{Myers2003}
{Myers}, S.~T., {Jackson}, N.~J., {Browne}, I.~W.~A., {de Bruyn}, A.~G., {Pearson}, T.~J., {Readhead}, A.~C.~S., et al. \href{http://adsabs.harvard.edu/abs/2003MNRAS.341....1M}{2003}, MNRAS, 341, 1

\bibitem[\protect\citeauthoryear{Narayan \& White}{1998}]{NarayanWhite1988}
{Narayan}, R., {White}, S.~D.~M., \href{http://adsabs.harvard.edu/abs/1988MNRAS.231P..97N}{1998}, MNRAS, 231, 97

\bibitem[\protect\citeauthoryear{Navarro et al.}{1996}]{Navarro1996}
{Navarro}, J.~F., {Frenk}, C.~S., {White}, S.~D.~M., \href{http://adsabs.harvard.edu/abs/1996ApJ...462..563N}{1996}, ApJ, 490, 493

\bibitem[\protect\citeauthoryear{Navarro et al.}{1997}]{Navarro1997}
{Navarro}, J.~F., {Frenk}, C.~S., {White}, S.~D.~M., \href{http://adsabs.harvard.edu/abs/1997ApJ...490..493N}{1997}, ApJ, 490, 493

\bibitem[\protect\citeauthoryear{Nayyeri et al.}{2016}]{Nayyeri2016}
{Nayyeri}, H., {Keele}, M., {Cooray}, A., {Riechers}, D.~A., {Ivison}, R.~J., {Harris}, A.~I., {Frayer}, D.~T., et al. \href{http://adsabs.harvard.edu/abs/2016ApJ...823...17N}{2016}, ApJ, 823, 17

\bibitem[\protect\citeauthoryear{Negrello et al.}{2007}]{Negrello2007}
{Negrello}, M., {Perrotta}, F., {Gonz{\'a}lez-Nuevo}, J., {Silva}, L., {de Zotti}, G., {Granato}, G.~L., et al. \href{http://adsabs.harvard.edu/abs/2007MNRAS.377.1557N}{2007}, MNRAS, 377, 1557

\bibitem[\protect\citeauthoryear{Negrello et al.}{2010}]{Negrello2010}
{Negrello}, M., {Hopwood}, R., {De Zotti}, G., {Cooray}, A., {Verma}, A., {Bock}, J., {Frayer}, D.~T., et al. \href{http://adsabs.harvard.edu/abs/2010Sci...330..800N}{2010}, Science, 330, 800

\bibitem[\protect\citeauthoryear{Negrello et al.}{2014}]{Negrello2014}
{Negrello}, M., {Hopwood}, R., {Dye}, S., {da Cunha}, E., {Serjeant}, S., {Fritz}, J., et al. \href{http://adsabs.harvard.edu/abs/2014MNRAS.440.1999N}{2014}, MNRAS, 440, 1999

\bibitem[\protect\citeauthoryear{Negrello et al.}{2017}]{Negrello2017}
{Negrello}, M., {Amber}, S., {Amvrosiadis}, A., {Cai}, Z.-Y., {Lapi}, A., {Gonzalez-Nuevo}, J., {De Zotti}, G., et al. \href{http://adsabs.harvard.edu/abs/2017MNRAS.465.3558N}{2017}, MNRAS, 465, 3558

\bibitem[\protect\citeauthoryear{Oguri}{2002}]{Oguri2002b}
{Oguri}, M., \href{http://adsabs.harvard.edu/abs/2002ApJ...580....2O}{2002}, ApJ, 580, 2

\bibitem[\protect\citeauthoryear{Oguri}{2006a}]{Oguri2006a}
{Oguri}, M., {Inada}, N., {Pindor}, B., {Strauss}, M.~A., {Richards}, G.~T., \href{http://adsabs.harvard.edu/abs/2006AJ....132..999O}{2006a}, MNRAS, 132, 999

\bibitem[\protect\citeauthoryear{Oguri}{2006b}]{Oguri2006b}
{Oguri}, M., \href{http://adsabs.harvard.edu/abs/2006MNRAS.367.1241O}{2006b}, MNRAS, 367, 1241

\bibitem[\protect\citeauthoryear{Oguri et al.}{2008}]{Oguri2008}
{Oguri}, M., {Inada}, N. {Strauss}, M.~A. {Kochanek}, C.~S. {Richards}, G.~T., et al., \href{http://adsabs.harvard.edu/abs/2008AJ....135..512O}{2008}, AJ, 135, 512

\bibitem[\protect\citeauthoryear{Oguri et al.}{2012}]{Oguri2012}
{Oguri}, M. {Inada}, N. {Strauss}, M.~A. {Kochanek}, C.~S. {Kayo}, I., et al., \href{http://adsabs.harvard.edu/abs/2012AJ....143..120O}{2012}, AJ, 143, 120

\bibitem[\protect\citeauthoryear{Oliver et al.}{2012}]{Oliver2012}
{Oliver}, S.~J. {Bock}, J. {Altieri}, B. {Amblard}, A. {Arumugam}, V. {Aussel}, H. {Babbedge}, T., et al., \href{http://adsabs.harvard.edu/abs/2012MNRAS.424.1614O}{2012}, MNRAS, 424, 1614

\bibitem[\protect\citeauthoryear{Omont et al.}{2013}]{Omont2013}
{Omont}, A. {Yang}, C. {Cox}, P. {Neri}, R. {Beelen}, A. {Bussmann}, R.~S. {Gavazzi}, R. {van der Werf}, P., et al., \href{http://adsabs.harvard.edu/abs/2013A\%26A...551A.115O}{2013}, A\&A, 551, A115

\bibitem[\protect\citeauthoryear{Perrotta et al.}{2002}]{Perrotta2002}
{Perrotta}, F. {Baccigalupi}, C. {Bartelmann}, M. {De Zotti}, G. {Granato}, G.~L., et al. \href{http://adsabs.harvard.edu/abs/2002MNRAS.329..445P}{2002}, MNRAS, 329, 445

\bibitem[\protect\citeauthoryear{Perrotta et al.}{2003}]{Perrotta2003}
{Perrotta}, F. {Magliocchetti}, M. {Baccigalupi}, C. {Bartelmann}, M. {De Zotti}, G. {Granato}, G.~L., et al., \href{http://adsabs.harvard.edu/abs/2003MNRAS.338..623P}{2003}, MNRAS, 338, 623

\bibitem[\protect\citeauthoryear{Pilbratt et al.}{2010}]{Pilbratt2010}
{Pilbratt}, G.~L. {Riedinger}, J.~R. {Passvogel}, T. {Crone}, G., et al., \href{http://adsabs.harvard.edu/abs/2010A\%26A...518L...1P}{2010}, A\&A, 518, 1

\bibitem[\protect\citeauthoryear{Planck Collaboration}{2014}]{Planck2014}
{Planck Collaboration} et al., \href{http://adsabs.harvard.edu/abs/2014A\%26A...571A..16P}{2014}, A\&A, 571, 16

\bibitem[\protect\citeauthoryear{Porciani \& Madau}{2000}]{PorcianiMadau2000}
{Porciani}, C. {Madau}, P., \href{http://adsabs.harvard.edu/abs/2000ApJ...532..679P}{2000}, ApJ, 532, 679

\bibitem[\protect\citeauthoryear{Prada et al.}{2012}]{Prada2012}
{Prada}, F. {Klypin}, A.~A. {Cuesta}, A.~J. {Betancort-Rijo}, J.~E., et al., arXiv:\href{http://adsabs.harvard.edu/abs/2012MNRAS.423.3018P}{2012}, MNRAS, 423, 3018

\bibitem[\protect\citeauthoryear{Prugniel \& Simien}{1997}]{PrugnielSimien1997}
{Prugniel}, P. {Simien}, F., \href{http://adsabs.harvard.edu/abs/1997A\%26A...321..111P}{1997}, A\&A, 321, 111

\bibitem[\protect\citeauthoryear{Schaller et al.}{2015a}]{Schaller2015a}
{Schaller}, M. {Frenk}, C.~S. {Bower}, R.~G. {Theuns}, T. {Jenkins}, A., et al., \href{http://adsabs.harvard.edu/abs/2015MNRAS.451.1247S}{2015}, MNRAS, 451, 1247

\bibitem[\protect\citeauthoryear{Schaller et al.}{2015b}]{Schaller2015b}
{Schaller}, M. {Frenk}, C.~S. {Bower}, R.~G. {Theuns}, T. {Trayford}, J., et al., \href{http://adsabs.harvard.edu/abs/2015MNRAS.452..343S}{2015}, MNRAS, 452, 343

\bibitem[\protect\citeauthoryear{Schaye et al. }{2015}]{Schaye2015}
{Schaye}, J. {Crain}, R.~A. {Bower}, R.~G. {Furlong}, M. {Schaller}, M., et al., \href{http://adsabs.harvard.edu/abs/2015MNRAS.446..521S}{2015}, MNRAS, 446, 521

\bibitem[\protect\citeauthoryear{Schneider, Ehlers \& Falco}{1992}]{SchneiderEhlersFalco1992}
{Schneider}, P. {Ehlers}, J. {Falco}, E.~E., \href{http://adsabs.harvard.edu/abs/1992grle.book.....S}{1992}, Gravitational Lenses, XIV (Berlin:Springer)

\bibitem[\protect\citeauthoryear{Sheth \& Tormen.}{1999}]{ShethTormen1999}
{Sheth}, R.~K. {Tormen}, G., \href{http://adsabs.harvard.edu/abs/1999MNRAS.308..119S}{1999}, MNRAS, 308, 199

\bibitem[\protect\citeauthoryear{Stanford et al. }{2014}]{Stanford2014}
{Stanford}, S.~A. {Gonzalez}, A.~H. {Brodwin}, M. {Gettings}, D.~P., et al., \href{http://adsabs.harvard.edu/abs/2014ApJS..213...25S}{2014}, ApJ, 213, 25

\bibitem[\protect\citeauthoryear{Takahashi \& Chiba}{2001}]{TakahashiChiba2001}
{Takahashi}, R. {Chiba}, T., \href{http://adsabs.harvard.edu/abs/2001ApJ...563..489T}{2001}, ApJ, 563, 489

\bibitem[\protect\citeauthoryear{Tinker et al.}{2008}]{Tinker2008}
{Tinker}, J. {Kravtsov}, A.~V. {Klypin}, A. {Abazajian}, K. {Warren}, M., et. al. \href{http://adsabs.harvard.edu/abs/2008ApJ...688..709T}{2008}, ApJ, 688, 709

\bibitem[\protect\citeauthoryear{Turner, Ostriker \& Gott}{1984}]{TurnerOstrikerGott1984}
{Turner}, E.~L. {Ostriker}, J.~P. {Gott}, III, J.~R., \href{http://adsabs.harvard.edu/abs/1984ApJ...284....1T}{1984}, ApJ, 284, 1

\bibitem[\protect\citeauthoryear{Viero et al.}{2014}]{Viero2014}
{Viero}, M.~P. {Asboth}, V. {Roseboom}, I.~G. {Moncelsi}, L. {Marsden}, G. {Mentuch Cooper}, E., et al., \href{http://adsabs.harvard.edu/abs/2014ApJS..210...22V}{2014}, ApJ, 210, 2

\bibitem[\protect\citeauthoryear{Wardlow et al.}{2013}]{Wardlow2013}
{Wardlow}, J.~L. {Cooray}, A. {De Bernardis}, F. {Amblard}, A. {Arumugam}, V. {Aussel}, H., et. al., \href{http://adsabs.harvard.edu/abs/2013ApJ...762...59W}{2013}, ApJ, 762, 59

\bibitem[\protect\citeauthoryear{White \& Rees}{1978}]{WhiteRees1978}
{White}, S.~D.~M. {Rees}, M.~J., \href{http://adsabs.harvard.edu/abs/1978MNRAS.183..341W}{1978}, MNRAS, 183, 341

\end{thebibliography}
\end{document}